%% file: dune_EM.tex
\documentclass[prl,aps,twocolumn,tightenlines,notitlepage,nofootinbib,preprintnumbers,letterpaper,superscriptaddress]{revtex4-1} 
\pdfoutput=1
\usepackage{epsfig}
\usepackage{amsfonts}
\usepackage{amsmath}
\usepackage{graphicx}
\usepackage{color}
\usepackage{amssymb,amsmath,hyperref,slashed,color,graphicx,bm}
\usepackage[T1]{fontenc}
\usepackage{relsize}
\usepackage{marvosym}
\usepackage{booktabs} 
\usepackage[normalem]{ulem}

\usepackage{accents}

\hypersetup{colorlinks,citecolor= nicegreen,linkcolor= nicered}
\definecolor{nicered}{rgb}{0.7,0.1,0.1}
\definecolor{nicegreen}{rgb}{0.1,0.5,0.1}

\graphicspath{{Figures/}}

\begin{document}

\preprint{FERMILAB-PUB-19-623-T, NUHEP-TH/19-17}

\title{Measuring the Weak Mixing Angle in the  DUNE Near Detector Complex}

\author{Andr\'{e} de Gouv\^{e}a}
\affiliation{Department of Physics \& Astronomy, Northwestern University, Evanston, IL 60208, USA}
\author{Pedro A.~N. Machado}
\affiliation{Theoretical Physics Department, Fermilab, P.O. Box 500, Batavia, IL 60510, USA}
\author{Yuber F. Perez-Gonzalez}
\affiliation{Department of Physics \& Astronomy, Northwestern University, Evanston, IL 60208, USA}
\affiliation{Theoretical Physics Department, Fermilab, P.O. Box 500, Batavia, IL 60510, USA}
\affiliation{Colegio de F\'isica Fundamental e Interdisciplinaria de las Am\'ericas (COFI), 254 Norzagaray street, San Juan, Puerto Rico 00901} 
\author{Zahra Tabrizi}
\affiliation{Instituto de F\'isica Gleb Wataghin, Universidade Estadual de Campinas (UNICAMP), Rua S\'ergio Buarque de Holanda, 777, Campinas, SP, 13083-859, Brazil}


\begin{abstract}
The planned DUNE experiment will have excellent sensitivity to the vector and axial couplings of the electron to the $Z$-boson via precision measurements of neutrino--electron scattering. We investigate the sensitivity of DUNE-PRISM, a movable near detector in the direction perpendicular to the beam line, and find that it will qualitatively impact our ability to constrain the weak couplings of the electron. We translate these neutrino--electron scattering measurements into a determination of the weak mixing angle at low scales and estimate that, with seven  years of data taking, the DUNE near-detector can be used to measure $\sin^2\theta_W$ with about 2\% precision.  We also discuss the impact of combining neutrino--electron scattering data with neutrino trident production at DUNE-PRISM. 
\end{abstract}

\pacs{Valid PACS appear here}

\maketitle

The standard model of particle physics (SM) is a quantum field theory with a $SU(3)_c\times SU(2)_L\times U(1)_Y$ gauge symmetry, corresponding to the color, weak-isospin and hypercharge interactions, respectively, along with a set of fermion and boson fields describing the particles observed in nature. Free SM parameters -- the gauge and Yukawa couplings, together with the scalar potential parameters -- need to be determined by comparing the results of theoretical computations to a finite set of the experimental measurements. 

The weak mixing angle $\theta_W$ (or, more precisely, its sine-squared, $\sin^2\theta_W$) parameterizes several measurable quantities: the mass ratio of the weak gauge bosons, some weak-interaction cross sections, and parity-violating observables. It is a crucial ingredient of the \emph{electroweak precision observables}, a set of experimental observables designed to test the SM internal consistency. 

The exact definition of the weak mixing angle depends on the renormalization scheme, that is, the convention of which quantities are taken as input and which are derived from these inputs, along with the recipe for handling quantum corrections. As quantum corrections are relevant, $\sin^2\theta_W$ depends on the scale at which it is being measured. For example, in the modified minimal subtraction scheme~\cite{Erler:2004in, Erler:2017knj}, $\overline{\scalebox{0.8}{\text{MS}}}$, the weak mixing angle is 
\begin{equation}
    \sin^2\theta_W(\mu) \equiv \frac{g^{\prime 2}(\mu)}{g^{2}(\mu)+g^{\prime 2}(\mu)},
\end{equation}
where $g$ and $g'$ are the $SU(2)_L$ and $U(1)_Y$ gauge coupling constants, respectively, and $\mu$ is the scale of the physical process under consideration. The SM predicts, under a specific renormalization scheme, a unique scale dependence for $\sin^2\theta_W$. This dependence has been confirmed by precise measurements at very different energy scales, including atomic parity violation, electron-proton scattering, M\"{o}ller scattering, neutrino--nucleus and neutrino--electron scattering, electron deep-inelastic scattering, and the $Z$- and $W$-boson masses (see Ref.~\cite{Tanabashi:2018oca} for a comprehensive review).

The NuTeV result~\cite{Zeller:2001hh}, the most precise measurement of $\sin^2\theta_W$ using neutrino scattering, stands out from the other measurements. Considering the ratios of neutral-current to charged-current and neutrino--nucleus to antineutrino--nucleus cross sections, they find $\sin^2\theta_W=0.2407\pm0.0016$ (in the $\overline{\scalebox{0.8}{\text{MS}}}$ scheme) at an average energy scale $\langle\mu\rangle \simeq 4.5$~GeV. This measurement deviates from the SM expectation anchored by the more precise measurements at LEP~\cite{ALEPH:2005ab} at the $3\sigma$ level. Effects inherent to the intricacies of neutrino-nucleus scattering, potentially unaccounted for or only partially accounted for by the collaboration, have been identified as candidate sources for the discrepancy~\cite{Pumplin:2002vw, Kretzer:2003wy, Sather:1991je, Rodionov:1994cg, Martin:2003sk, Londergan:2003ij, Bentz:2009yy, Gluck:2005xh, Kumano:2002ra, Kulagin:2003wz, Brodsky:2004qa, Hirai:2004ba, Miller:2002xh, Cloet:2009qs, Diener:2003ss, Arbuzov:2004zr, Park:2009ft, Diener:2005me, Dobrescu:2003ta}. A definitive answer remains elusive. Regardless, it stands to reason that other precise measurements of $\sin^2\theta_W$ using neutrino scattering will help shed light on the situation.

Next-generation neutrino experiments like LBNF-DUNE~\cite{Acciarri:2015uup} and T2HK~\cite{Abe:2011ts} will include very intense neutrino beams with energies that range from several hundred MeV to several GeV. The neutrino--nucleus scattering cross sections, at these energies, have large uncertainties due to nuclear and non-perturbative effects~\cite{Alvarez-Ruso:2017oui}, making it very challenging to use them to infer $\sin^2\theta_W$. Neutrino--electron scattering, on the other hand, provides a more promising environment \cite{Conrad:2004gw,deGouvea:2006hfo,Agarwalla:2010ty,Conrad:2013sqa,Adelmann:2013isa}. Even in this case, however, one still needs to address significant challenges. First, the cross section for neutrino--electron scattering is three orders of magnitude smaller than that for neutrino-nucleus scattering, translating into poor statistics in most neutrino experiments. Second, while the neutrino--electron cross section depends mostly on $\sin^2\theta_W$, the neutrino beam originates from the in-flight decay of charged mesons produced by high-energy protons hitting a fixed target. First-principles computations of the meson production rate and kinematics are not possible and one must rely on phenomenological models and experimental data; uncertainties on the overall neutrino flux and energy distribution are at the $5\%$ to $15\%$ level~\cite{Abe:2012av, Aliaga:2016oaz, Marshall:2019vdy}.

Near detector complexes are designed to circumvent some of the large uncertainties in the flux and cross sections and allow precision measurements of neutrino oscillations~\cite{Marshall:2019vdy}. DUNE-PRISM \cite{pickering-prism}, currently part of the LBNF-DUNE proposal, is a near detector that is capable of moving in the direction perpendicular to the neutrino-beam axis.  Although the neutrino flux has prohibitively large uncertainties, the ratios of on-axis to off-axis fluxes are dictated only by meson-decay kinematics and thus are much better understood. Therefore, measurements of the neutrino-electron-scattering spectrum at different off-axis positions should allow an unprecedented measurement of the weak mixing angle with neutrinos.

In general terms, the neutrino--electron scattering cross-section depends on the vector and axial couplings, $g_V$ and $g_A$, between the $Z$-boson and the electron (see CHARM-II~\cite{Vilain:1994qy}, LSND~\cite{Auerbach:2001wg} and TEXONO~\cite{Deniz:2009mu}). We will estimate the DUNE-PRISM sensitivity to such parameters via neutrino-electron scattering data. A hidden but very safe assumption is that the cross-section depends only on the neutrino left-handed coupling to the $Z$-boson. The reason for this is that all neutrinos and antineutrinos used in neutrino scattering are produced in charged-current processes ($\pi^+\to\mu^+\nu_{\mu}$, $n\to p \,e\,\bar{\nu}_e$, $D_s\to\tau^+\nu_{\tau}$, etc) and are, to a very good precision, 100\% polarized. Lepton-collider data, combined with those from neutrino--electron scattering, for example, can be used to determine the right-handed-coupling of neutrinos to the $Z$-boson \cite{Carena:2003aj}. 

The differential cross section for a neutrino with flavor $\alpha=e,\mu,\tau$ to scatter off an electron at rest is
\begin{align}
    	\frac{d\sigma}{dE_R}  &= \frac{2G_{F}^2m_e}{\pi}\left\{g_1^2+g_2^2\left(1-\frac{E_R}{E_\nu}\right)^2-g_1 g_2\frac{m_e E_R}{E_\nu^2}\right\} \nonumber \\ \label{eq:xsec}
	&\simeq1.72\times10^{-41}\left\{g_1^2+g_2^2\left(1-\frac{E_R}{E_\nu}\right)^2\right\} \frac{{\rm cm}^2}{\rm GeV},
\end{align}
where $G_F$ is the Fermi constant, $E_\nu$ is the incoming neutrino energy, and $m_e$ and $E_R$ are the electron mass and recoil kinetic energy, respectively. 

The couplings $g_1$ and $g_2$ depend on the neutrino flavor and can be written in terms of  $g_V$ and $g_A$; thus, they can be expressed in terms of $\sin^2\theta_W$, see Table~\ref{tab:couplings}. More generally, if $g_V$ and $g_A$ are considered to be free parameters, they can be independently extracted from the recoil-electron energy spectrum so data from DUNE-PRISM are expected to constrain nontrivial regions in the $g_V\times g_A$ plane. 

\begin{table}[ht]
\caption{Couplings $g_1$ and $g_2$ (see Eq.~\eqref{eq:xsec}) as a function of the electron--$Z$-boson couplings $g_V$ and $g_A$, for each neutrino flavor, along with the corresponding SM value. $s^2_W\equiv\sin^2\theta_W$.  \label{tab:couplings}}
    \centering
    \begin{tabular}{lcccc}
        \toprule\toprule
        $\nu_\alpha$ & $g_1$ & $g_1$(SM) & $g_2$ & $g_2$(SM) \\ \midrule\midrule
        $\nu_e$ & $1\!+\!(g_V\!+\!g_A)/2$ & $1/2\!+\!s_W^2$ & $(g_V\!-\!g_A)/2$& $s_W^2$ \\ \midrule
        $\nu_{\mu,\tau}$ & $(g_V\!+\!g_A)/2$ & $-1/2\!+\!s^2_W$ & $(g_V\!-\!g_A)/2$& $s^2_W$\\ \midrule
        $\bar\nu_{e}$ & $(g_V\!-\!g_A)/2$ & $s^2_W$ & $1\!+\!(g_V+g_A)/2$& $1/2\!+\!s_W^2$\\  \midrule
        $\bar\nu_{\mu,\tau}$ & $(g_V\!-\!g_A)/2$ & $s^2_W$ & $(g_V\!+\!g_A)/2$ & $-1/2\!+\!s^2_W$\\  \bottomrule
    \end{tabular}
\end{table}

Strictly speaking, the neutrino--electron cross section is also subject to quantum corrections that will introduce additional dependence on $Q^2\equiv 2E_R m_e$ \cite{Tomalak:2019ibg,Hill:2019xqk}. Kinematics dictates that the maximum recoil energy is approximately $E_R^{\rm max}\simeq E_\nu-m_e/2$. Due to kinematics and the energy profile of DUNE's neutrino flux, most electron recoil events will lie within $0.2\lesssim E_R\lesssim 10$~GeV. Therefore, the $Q^2$ values accessible to DUNE are, roughly, in the range $(10-100~{\rm MeV})^2$, where loop corrections to $\sin^2\theta_W$ have little scale dependence~\cite{Tanabashi:2018oca}\footnote{We have checked numerically that flavor-dependent electroweak corrections do not change our results at the 1$\sigma$ level.}. Thus, by analyzing the $Q^2$ distribution in detail, the couplings in Eq.~(\ref{eq:xsec}) can be interpreted as the renormalized couplings in the $\overline{\scalebox{0.8}{\text{MS}}}$ scheme at an average scale $\langle Q^2\rangle = (55~{\rm MeV})^2$.

Assuming the SM, the cross section for $\nu_\mu-e$-scattering is 
\begin{equation}
\frac{d\sigma}{dE_R}\propto \left(\frac{1}{4}-\sin^2\theta_W\right)+\sin^4\theta_W\left(2-\frac{2E_R}{E_\nu}+\frac{E_R^2}{E_\nu^2}\right).
\end{equation}
Since $\sin^2\theta_W$ is close to 1/4, the first term is suppressed relative to the second one. This implies that the value of $\sin^2\theta_W$, to leading order, modifies the overall normalization of the $\nu_\mu-e$-scattering cross section and the effect of changing $\sin^2\theta_W$ is nearly degenerate with that of changing the overall normalization of the $\nu_{\mu}$ flux. The situation is different for $\nu_e-e$ and  $\bar{\nu}_e-e$ scattering; $\sin^2\theta_W$ has a significant impact on the shape of the recoil-electron energy distributions. It turns out, unfortunately, that, at DUNE, the neutrino flux is dominated by $\nu_{\mu}$ and the $\nu_e$ contribution is relatively small, around a few percent. 

In this context, DUNE-PRISM is expected to provide nontrivial information. In accelerator neutrino experiments, the $\nu_{\mu}$ comes predominantly from the two-body decay $\pi^+\to\mu^+ \nu_\mu$ (and $K^+\to\mu^+\nu_\mu$, to a lesser extent) while the $\nu_e$ comes from the three-body decays of kaons and muons. For the same parent energy, the flux of $\nu_e$ has a larger angular spread than that of $\nu_{\mu}$ so the off-axis $\nu_e$ to $\nu_{\mu}$ flux ratio is larger than the on-axis one. 

\begin{figure}[ht]
  \centering
  \includegraphics[width=0.9\linewidth]{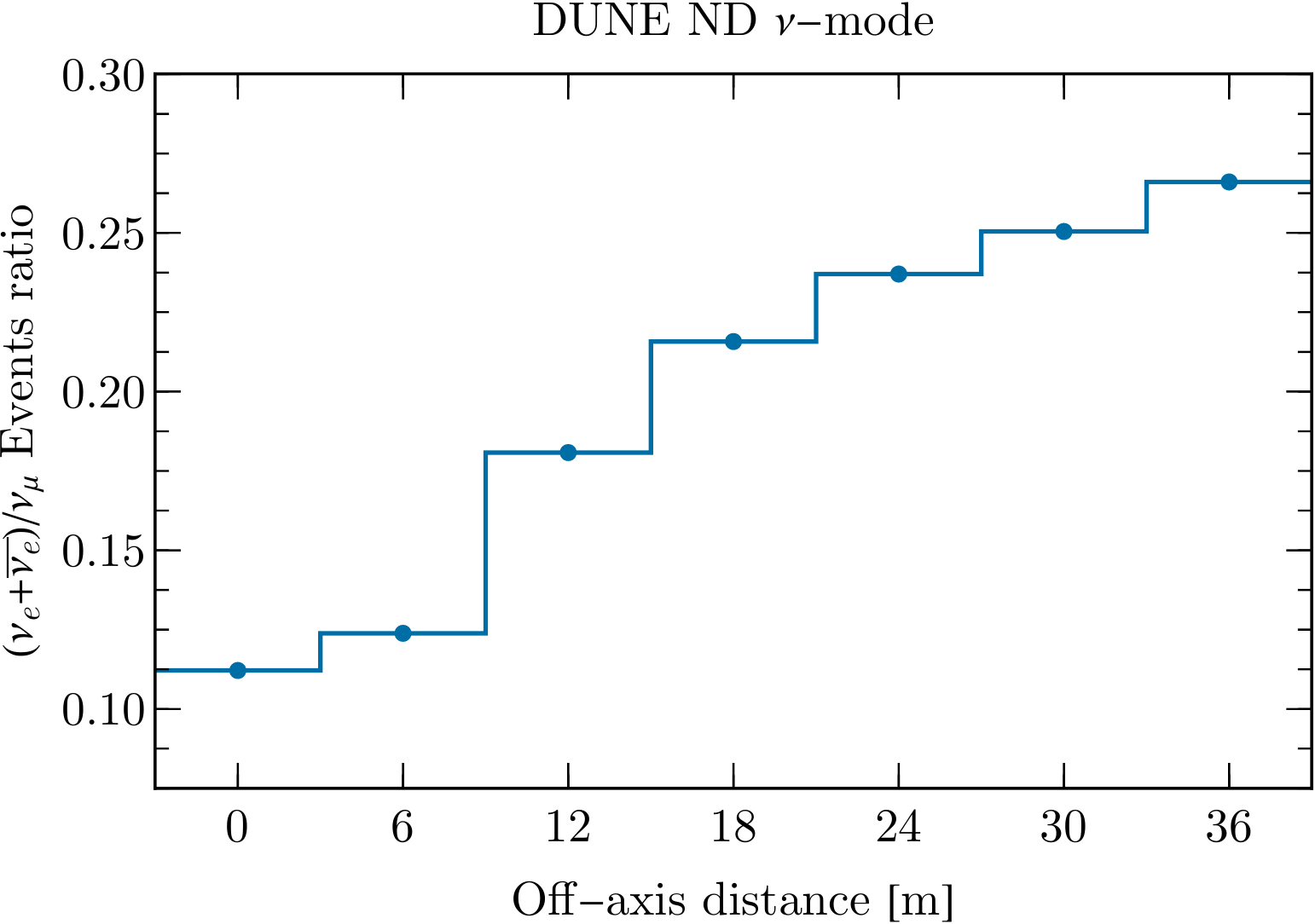}
  \caption{Ratio of the number of events expected from $\nu_e-e+\bar{\nu}_e-e$-scattering to that of $\nu_{\mu}-e$-scattering, in neutrino-mode running, as a function of the off-axis distance. 
  \label{fig:ratio}}
\end{figure}

To estimate how well DUNE-PRISM can contribute to the precision electroweak physics program, we compute the sensitivity to $\sin^2\theta_W$ for both on-axis and off-axis runnings. For concreteness, we assume seven years of data taking equally divided between neutrino and antineutrino modes. We assume a 75~ton fiducial mass liquid argon time projection chamber (LArTPC) and a 1.2~MW proton beam, as described in the DUNE Conceptual Design Report~\cite{Acciarri:2015uup}. 
For the off-axis configuration, we assume the near detector will take data at seven different positions, half of the time on-axis and half of the time equally divided in the off-axis positions. The detector is assumed to be 574~m away from the source in the beam axis direction while its transverse distances to the beam axis are $6N$~meters, $N=0,\ldots 6$. The detector experiences at each position a flux that is approximately $10N$~mrad off-axis, respectively. Fig.~\ref{fig:ratio} depicts the ratio of the number of events expected from $\nu_e-e$ and $\bar{\nu}_e-e$-scattering to that of $\nu_{\mu}-e$-scattering, in neutrino-mode running, as a function of the off-axis distance. As expected, the relevance of the $\nu_e$ and $\bar{\nu}_e$-initiated events grows significantly with the off-axis angle. Note that, while the flux ratio is of order a few percent, the $\nu_e-e$-scattering cross section is larger than the $\nu_\mu-e$ so, even on-axis, the $\nu_e$ contribution is of order 10\%.

To account for the energy-dependent neutrino-flux uncertainties and the correlations between the fluxes at different off-axis angles, we make use of a covariance matrix spanning all DUNE-PRISM positions and neutrino flavors, derived from detailed simulations of hadron production in the beam target followed by magnetic-horn focusing of charged particles~\cite{privatecomm}. The binning is performed in $E_e\theta_e^2$, where $E_e=E_R+m_e$ is the total electron energy and $\theta_e$ is the electron scattering angle relative to the beam direction (see Supplemental Material for details.). We consider a threshold kinetic energy $E_R>50~$MeV and perform the analysis in the range $(0.05<E_R<20)$~GeV. 

The main backgrounds for neutrino--electron scattering are charged-current quasi-elastic (CCQE) $\nu_e$-scattering events, $\nu_e A \to e^- A'$ and mis-identified $\pi^0$ events with no detectable hadronic activity, $\nu A\to \nu \pi^0 A$. 
Although the $\nu_e$ flux is only a few percent of the total neutrino flux, the CCQE cross section is over 1000 times larger than that for neutrino--electron scattering. 
We simulate these backgrounds using the NuWro event generator~\cite{Golan:2012wx}, and allow a 10\% normalization uncertainty of both of them. We cut any event which have at least one protons with kinetic energy above 50~MeV. For the $\pi^0$ background, we also require one photon to be soft, below 30~MeV, to be accepted.In principle, if the photons are sufficiently collinear, the two showers could be mis-identified as an electron event.
As the minimum photon-photon opening angle is $\theta> 15.5^\circ ({\rm GeV}/E_{\pi^0})$, it is unlikely that this poses a background and therefore we have neglected it.

Kinematics limit $E_e\theta_e^2<2m_e$ for  neutrino--electron scattering and thus we bin on $E_e\theta_e^2$ to improve background rejection.
LArTPCs have an exquisite angular resolution, of order $1^\circ$, for electromagnetic showers~\cite{Acciarri:2015uup}. 
In the Supplemental Material, we show how angular resolution affects the $E_e\theta_e^2$ spectrum and the sensitivity to $\sin^2\theta_W$.

\begin{figure}[t]
  \centering
  \includegraphics[width=\linewidth]{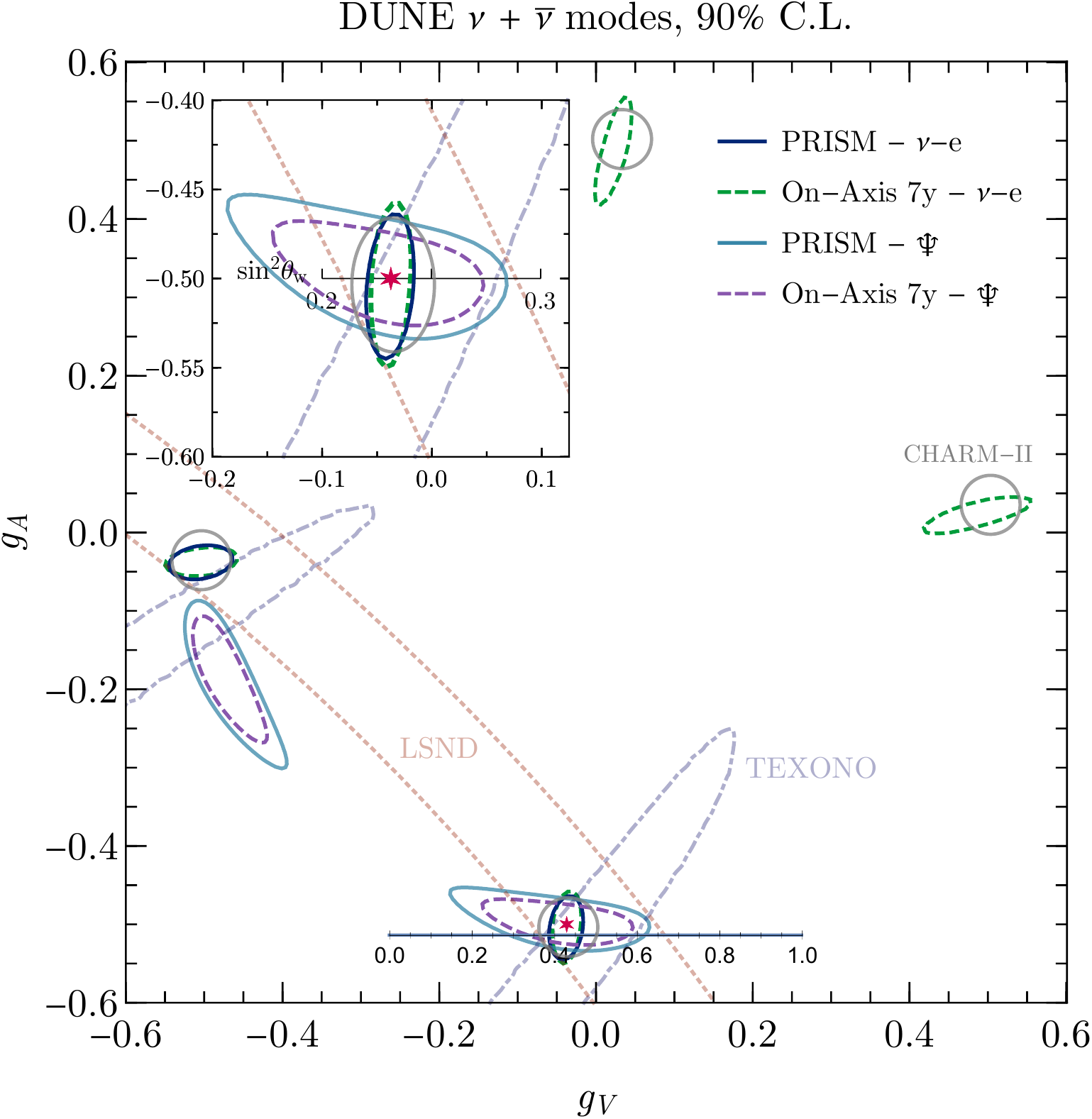}
  \caption{Allowed regions in the plane $g_V\times g_A$ from CHARM-II~\cite{Vilain:1994qy} (gray, at 90 \%C.L.), LSND~\cite{Auerbach:2001wg} (dotted light-brown, at 1$\sigma$ C.L.), and TEXONO~\cite{Deniz:2009mu} (dot-dashed light-violet, at 1$\sigma$ C.L.), and the estimated 90\% C.L.~sensitivity from on-axis (7 years) DUNE electron scattering (dashed green) and tridents (\Neptune) (dashed purple), as well as DUNE-PRISM electron scattering (dark-blue) and tridents (light-blue), assuming the SM value (red star).
\label{fig:gv-ga}}
\end{figure}

Fig.~\ref{fig:gv-ga} depicts the DUNE sensitivity to the vector and axial couplings, $g_V$ and $g_A$, in the on-axis LArTPC (dashed green) or the DUNE-PRISM configuration (dark-blue). For comparison, we include existing measurements from CHARM-II~\cite{Vilain:1994qy} (gray), LSND~\cite{Auerbach:2001wg} (dotted light-brown) and TEXONO~\cite{Deniz:2009mu} (dot-dashed light-violet). Both the DUNE on-axis and CHARM-II measurements suffer from a four-fold degeneracy; this is a consequence of the fact that the neutrino flux in both these experiments is dominated by $\nu_{\mu}$. There is an exact  degeneracy in the differential cross section for $\nu_\mu-e$ scattering under the transformations
\begin{equation}
    (g_V,g_A)\to(g_A,g_V)\,\,\,\text{ and }\,\,\, (g_V,g_A)\to(-g_V,-g_A),
\end{equation}
see Eq.~\eqref{eq:xsec} and Table~\ref{tab:couplings}, and hence an experiment with a pure $\nu_{\mu}$ beam is intrinsically limited. The TEXONO experiment measured electron recoils from electron anti-neutrinos produced in a nuclear reactor. The scattering cross section, in this case, is proportional to $3g_1^2+g_2^2$, which defines an oblique ellipse in the $(g_V,g_A)$ plane centered at $(-0.5,-0.5)$. The TEXONO result in Fig.~\ref{fig:gv-ga} reflects this fact, up to effects related to information on the recoil energy spectrum. The LSND measurement can also be understood by noticing that the flux of neutrinos consists of $\nu_\mu$, $\bar\nu_\mu$, and $\nu_e$ with well-characterized energy spectra from (mostly) pion decay at rest, followed by muon decay at rest. Current data are not able to rule out very small $g_A$ and $g_V\sim -0.5$ (region on the left-hand part of Fig.~\ref{fig:gv-ga})

\begin{figure}[t]
  \centering
  \includegraphics[width=\linewidth]{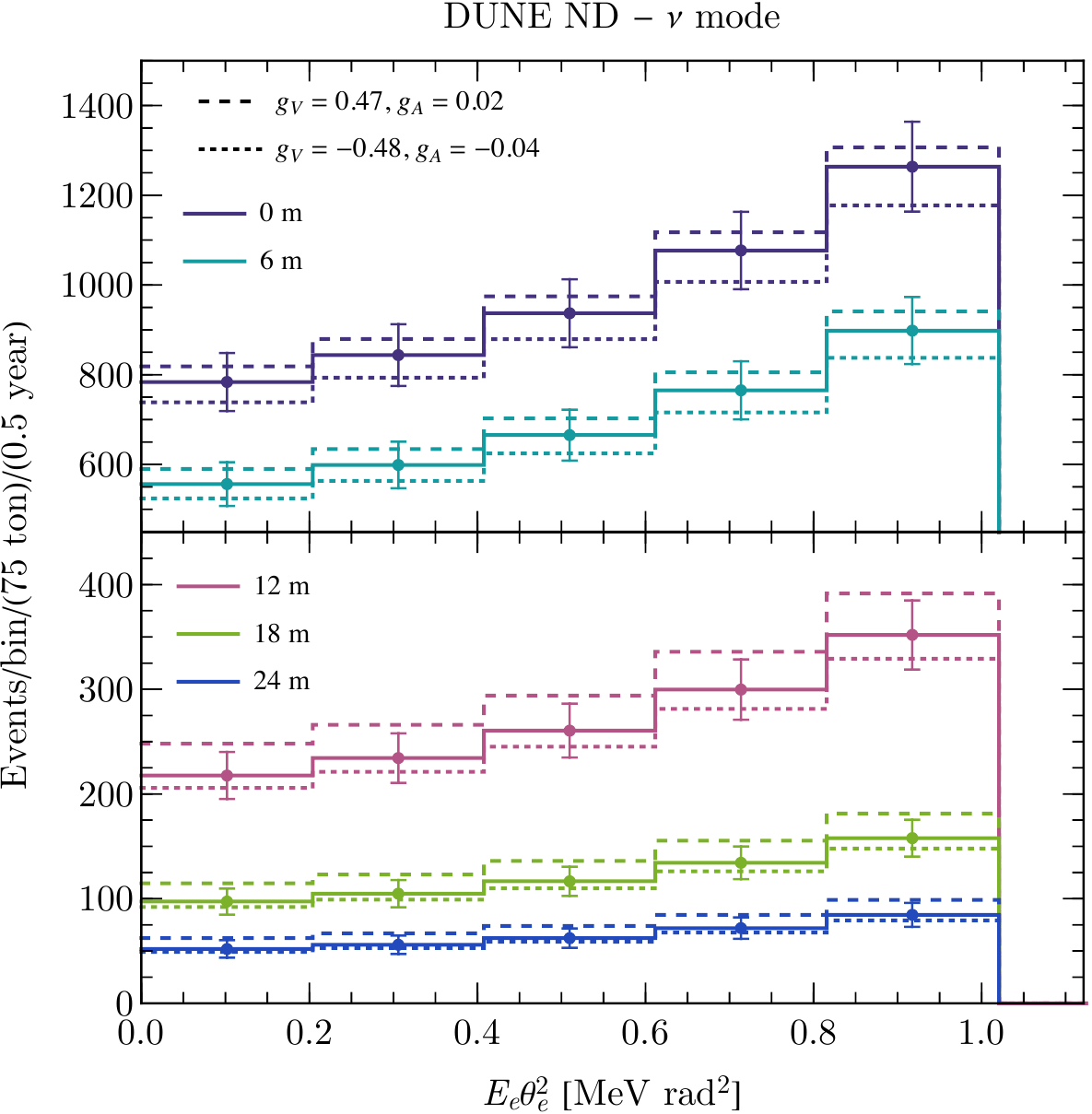}
  \caption{Neutrino-electron event rates as a function of $E_e\theta_e^2$ for the first 5 off-axis positions. For each position, the three histograms  correspond to three pairs of vector and axial couplings $(g_V,g_A)$: $(-0.04, -0.5)$ (solid); $(-0.48, -0.04)$ (dotted); and $(0.47, 0.02)$ (dashed). Error bars illustrating the statistical and systematic errors, are included for the SM case (solid histogram).
  \label{fig:spectra}}
\end{figure}

In DUNE-PRISM, the presence of both $\nu_{\mu}$ and $\nu_e$, along with their antiparticles, is a powerful tool for lifting degeneracies without resorting to data from other experiments. 
To illustrate this point, Fig.~\ref{fig:spectra} depicts the $E_e\theta_e^2$ spectra of neutrino-electron scattering events for the first 5 off-axis positions without any angular or energy resolution.  
For each position, histograms corresponding to three pairs of vector and axial couplings $(g_V,g_A)$ are depicted: $(-0.04, -0.5)$, the SM expectation (solid); $(-0.48, -0.04)$, the leftmost degenerate region (dotted); and $(0.47, 0.02)$, the rightmost degenerate region (dashed). 
It is clear that the rightmost degeneracy is lifted due to the higher $\nu_e$ composition of the flux, as depicted in Fig.~\ref{fig:ratio}. 
Error bars illustrating the statistical and systematic errors, are included for the SM case. 
DUNE-PRISM neutrino--electron scattering data, alone, cannot fully distinguish the SM from the leftmost degenerate region, as depicted in Fig.~\ref{fig:gv-ga}.

Neutrino-trident scattering, when a neutrino scatters off a nucleus producing a charged lepton pair with same or different flavors, $\nu A\to\nu \ell^{+}  \ell^{-} \! A$, is also sensitive to $g_V$ and $g_A$.\footnote{Here we assume that the electron and muon couplings to the $Z$-boson are identical.} This scattering can be  coherent off the electromagnetic field of the nucleus or diffractive off the nucleons themselves. Although the trident cross section is quite involved (see e.g. Refs~\cite{Magill:2016hgc, Ballett:2018uuc, Altmannshofer:2019zhy}), in the limit where the final state leptons are massless, it is proportional to the electroweak parameters $(C_V^2+C_A^2)$.  For a $\nu_\mu$ beam, these couplings are~\cite{Magill:2016hgc, Ballett:2018uuc}
\begin{align}\label{eq:trident-1}
    C_V  &= g_V  & C_A &= g_A \quad&(e^+e^- \,{\rm trident}),\\
    C_V  &= g_V+1 & C_A&= g_A+1 \quad &(\mu^+\mu^-\,{\rm trident}).\label{eq:trident-2}
\end{align}
The processes that lead to $\mu^{\mp}e^{\pm}$ tridents are pure-charged-current and do not contribute to this discussion. 
Hence, measurements of $e^+e^-$ $\nu_{\mu}$-tridents -- the statistically-dominant mode -- constrain $g_V^2+g_A^2$ while those of $\mu^+\mu^-$ $\nu_{\mu}$-tridents constrain $(g_V+1)^2+(g_A+1)^2$ in the limit of vanishing muon mass. 
A similar behavior is expected of $\nu_{e}$-tridents, with $e\leftrightarrow \mu$, and those associated to antineutrinos. It is easy to see that, in the limit where the muon mass vanishes, all cross sections are invariant under $g_V\leftrightarrow g_A$. A finite muon mass, however, breaks the $g_V\leftrightarrow g_A$ symmetry. 

Due to the very high intensity of the DUNE neutrino beam, this rare process is accessible. Fig.~\ref{fig:gv-ga} also depicts the measurement of $(g_V,g_A)$ from both $\mu^+\mu^-$ and $e^+e^-$ neutrino-trident events in DUNE on-axis (dashed purple) and DUNE-PRISM (light-blue), considering the efficiencies from \cite{Ballett:2018uuc}, which range from $48-66\%$ ($17-39\%$) for coherent (diffractive) trident processes. These efficiencies stem from cuts on hadronic activity and kinematical variables in order to make backgrounds negligible. Improvements on the reconstruction of di-electron or di-muon events would benefit the $(g_V,g_A)$ couplings determination. The allowed region is not symmetric under $g_V\leftrightarrow g_A$ since, as highlighted earlier, for DUNE energies, the mass of the muon is not negligible. Indeed, we checked that the subleading  $\mu^+\mu^-$ neutrino-trident event sample plays the decisive role here. Hence, the combination of neutrino--electron scattering and neutrino trident data in the DUNE near detector complex, assuming these are consistent with the SM,  lifts all degeneracies in the $g_V\times g_A$ plane even if one chooses to exclude information from outside data. 

\begin{figure}[t]
  \centering
  \includegraphics[width=\linewidth]{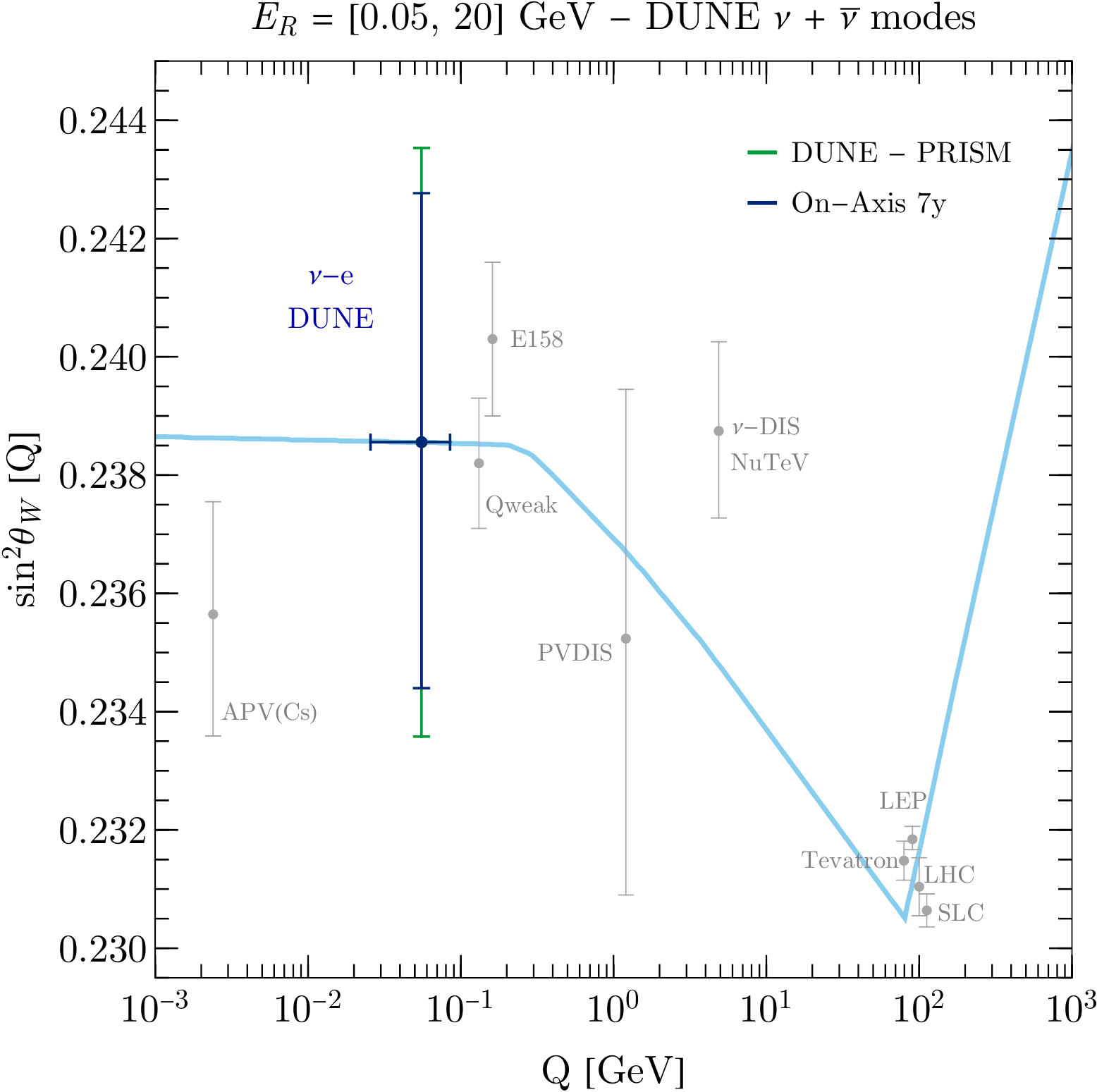}
  \caption{$\sin^2\theta_W$ in the $\overline{\rm MS}$ scheme (light blue line) as a function of $Q$, obtained from a fit to existing data (gray data points), together with the DUNE on-axis (dark blue data point) and DUNE-PRISM (green data point) sensitivities to this angle. The horizontal error bars indicate the range of $Q$ values accessible to DUNE neutrino--electron scattering. Note that the Tevatron, LHC and SLC data points where slightly shifted from $Q=M_Z$ to improve readability.  \label{fig:sin2qw}}
\end{figure}

Assuming the SM, our results can be translated into a measurement of $\sin^2\theta_W$ at $\langle Q^2\rangle = (55~{\rm MeV})^2$. Fig.~\ref{fig:sin2qw} depicts the value of $\sin^2\theta_W$ in the $\overline{\scalebox{0.8}{\text{MS}}}$ scheme as a function of $Q$, obtained from a fit to existing data, together with our estimate for the expected  DUNE and DUNE-PRISM sensitivities. The former is slightly better, but we emphasize that the on-axis measurement of $\sin^2\theta_W$ depends more strongly on the neutrino-flux modeling, while the DUNE-PRISM sensitivity depends more on the relative on- to off-axis flux uncertainties. The main systematic uncertainty for this analysis comes from hadron production in the beam target, and extra running time would further improve the determination of $\sin^2\theta_W$ (see Supplemental Material). Note that current experiments like NA61/SHINE~\cite{Abgrall:2011ae} or the future experiment EMPHATIC~\cite{Akaishi:2019dej} may achieve a better knowledge of the hadron production mechanism leading to reduced systematic uncertainties and thus improving the determination of the weak mixing angle. Regardless, both measurements are estimated to be competitive with existing results.

In summary, we estimated that the future DUNE experiment will have excellent sensitivity to the vector and axial couplings of the electron to the $Z$-boson, and thus to the weak mixing angle $\sin^2\theta_W$, via precision measurements of neutrino--electron scattering. The sub-dominant $\nu_e$ beam component in DUNE-PRISM, as well as neutrino trident events,  play an important role in resolving degeneracies currently present in the world data. 

\acknowledgments

We are extremely grateful to Luke Pickering for providing us with the flux covariance matrix, and we thank Laura Fields and Oleksandr Tomalak  for useful discussions. 
The work of AdG is supported in part by the DOE Office of Science award~\#DE-SC0010143.
This manuscript has been authored by Fermi Research Alliance, LLC under Contract No. DE-AC02-
07CH11359 with the U.S. Department of Energy, Office of Science, Office of High Energy Physics. ZT is supported by Funda\c{c}\~{a}o de Amparo \`{a} Pesquisa do Estado de S\~{a}o Paulo (FAPESP) under contract 2018/21745-8.

\bibliography{references}

\input{DUNE_tw_SM.tex}

\end{document}

%% file: DUNE_tw_SM.tex
\newpage
\onecolumngrid
\appendix
\section{Supplemental Material}


In this Supplemental Material, we provide technical details that may be relevant to experts.

\begin{figure}[t]
  \centering
  \includegraphics[width=\linewidth]{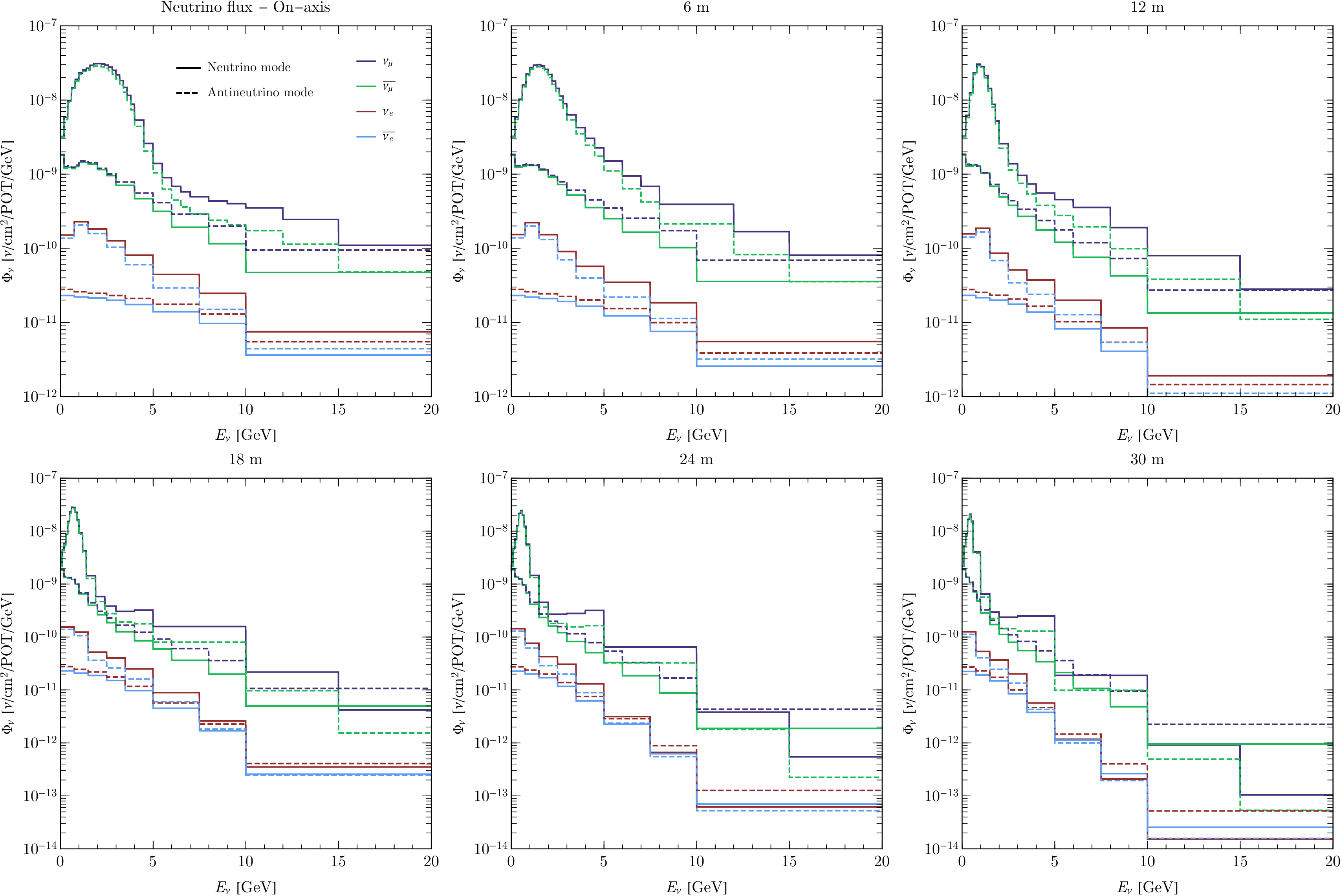}
  \caption{Neutrino fluxes considered in the on-axis and five off-axis positions. We show the fluxes for $\nu_\mu$ (purple), $\overline{\nu_\mu}$ (green), $\nu_e$ (red) and $\overline{\nu_e}$ (light blue) in neutrino (full) and antineutrino (dashed) running modes.}
  \label{fig:Flux}
\end{figure}

In Fig. \ref{fig:Flux}, we present the neutrino fluxes considered in the analysis at the near detector facility for the on-axis and 5 off-axis positions and for both running -- neutrino and antineutrino -- modes.
The number of expected neutrino-electron scattering events in the DUNE liquid argon near detector, for each neutrino flavor in each position, is shown in Tab.~\ref{tab:LArrates}.
The numbers in the top rows (bottom rows) correspond to neutrino (antineutrino) mode runs.
We have assumed 3.5 years of data taking in each mode, half of the time spent in the on-axis position, and the other half equally distributed in each off-axis position.

\begin{table}[h]
\begin{center}
\scalebox{0.9}{
\begin{tabular}{|c|c|c|c|c|c|c|c|}
\hline\hline
		\bf Channel & \bf  0 m& \bf 6 m & \bf 12 m & \bf 18 m &\bf  24 m &\bf  30 m &\bf  36 m  \\ \hline \hline
    $\nu_\mu e\to \nu_\mu e$& 13,624 & 1,576  & 565  & 238  &  122 & 72  & 47   \\
    & 2,025 & 269  & 129  & 65  & 37  &  25 &  18 \\\hline
    $\bar\nu_\mu e\to \bar\nu_\mu e$& 1,238  & 165  &  80 &  40 & 23  &  15 & 11  \\
    & 10,135  & 1,144  & 397  & 162  &  83 & 49  & 32  \\\hline
   $\nu_e e\to \nu_e e$ & 1,447 &  184 &  96 &  48 &  26  & 17  &  11 \\
    & 629  & 83  & 46  & 25  & 15  & 10  & 7  \\
    \hline
    $\bar\nu_e e\to \bar\nu_e e$ & 192  & 25  & 15  & 8  &  5  &  3 & 2  \\
    & 415  & 52  & 27  & 14  & 8   &  5  & 3   \\
    \hline\hline
    {\rm{Total}} & 16,501   &  1,950    &  756  &  334  &  176  & 107   & 71  \\
        & 13,204  & 1,548    &  599 &   266 &   143 &   89  &   60\\
        \hline\hline
\end{tabular}
}
\end{center}
\caption{\label{tab:LArrates} 
Total expected number of $\nu-e$ events at different off-axis positions for the neutrino mode (top row) and anti-neutrino mode (bottom row) assuming $3.5$ years of data taking in each mode. Half of the time is spent in the on-axis position while the other half is divided equally in each off-axis position.
 }
\end{table}

We also present in Tab.~\ref{tab:LArrates-tridents}, the expected number of trident events for the same running plan aforementioned.

\begin{table}[h]
\begin{center}
\scalebox{0.9}{
\begin{tabular}{|c|c|c|c|c|c|c|c|}
\hline\hline
		\bf Channel & \bf 0 m& \bf 6 m & \bf 12 m & \bf 18 m &\bf  24 m &\bf  30 m &\bf  36 m  \\ \hline \hline
    Total $e^\pm \mu^\mp$&  760  & 91  & 31  & 12  &  5 &  3 &  2  \\
    & 651  & 76  & 25  & 10  &  5 &  3 &  2 \\\hline
    Total $e^+ e^-$ &  180  & 21  & 8  & 3  &  2 &  1 &  0.6 \\
    & 166  & 19  & 7  & 3  &  1 &  1 &  0.6 \\\hline
   Total $\mu^+ \mu^-$ & 93  & 12  & 4  & 1  &  0.6  &  0.3 &  0.2 \\
    & 77  & 10  & 3  & 1  &  0.5 &  0.3 &  0.2 \\
    \hline\hline
\end{tabular}}
\end{center}
\caption{\label{tab:LArrates-tridents}Total expected number of trident events at different off-axis positions for the neutrino mode (top row) and anti-neutrino mode (bottom row) assuming $3.5$ years of data taking in each mode. Half of the time is spent in the on-axis position while the other half is divided equally in each off-axis position. }
\end{table}

The test statistics considered in the analysis is defined as, 
\begin{align}
   \chi^2 = (\mathbf{D}-\mathbf{T}-\alpha_{CC}\mathbf{B}^{\rm CCQE}-\alpha_{NC}\mathbf{B}^{\rm \pi^0})^T C^{-1}(\mathbf{D}-\mathbf{T}-\alpha_{CC}\mathbf{B}^{\rm CCQE}-\alpha_{NC}\mathbf{B}^{\rm \pi^0}) + \left(\frac{\alpha_{CC}}{\sigma_{\alpha}}\right)^2+\left(\frac{\alpha_{NC}}{\sigma_{\alpha}}\right)^2,
 \end{align}
where $\mathbf{D}$ and $\mathbf{T}$ are vectors of Asimov data and expectation values, respectively, spanning all positions and bins. The binning is performed in $E_e\theta_e^2$, $E_e$ the total electron energy and $\theta_e$ the scattered electron angle with respect to the beam direction. $C$ is the covariance matrix derived from detailed simulations of hadron production. The vectors $\mathbf{B}^{\rm CCQE}$ and $\mathbf{B}^{\rm \pi^0}$ correspond to the simulated background events from CCQE interactions and mis-identified  $\pi^0$ events, respectively. $\alpha_{CC}$ and $\alpha_{NC}$ are systematic uncertainties related to the normalization of such charged-current and neutral-current backgrounds. We include a penalty term for these systematic uncertainties with $\sigma_\alpha=10\%$

\begin{figure}[h]
  \centering
  \includegraphics[width=0.5\linewidth]{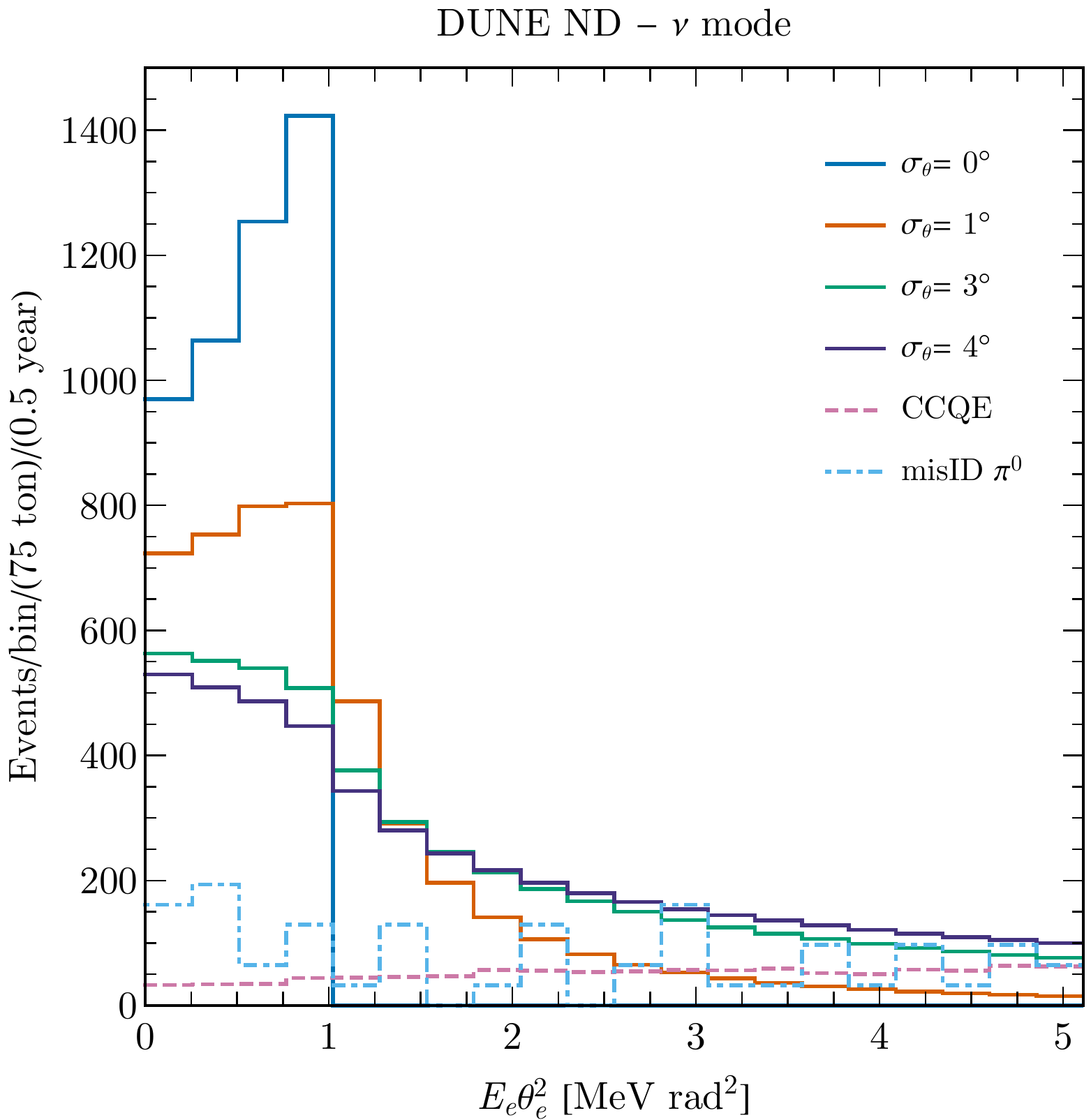}
  \caption{Neutrino-electron event rates as function of $E_e\theta_e^2$ for different values of the angular resolution. The pink dashed (light blue dot-dashed) lines correspond to the CCQE (misID $\pi^0$) estimated backgrounds.}
  \label{fig:Specsm}
\end{figure}

Regarding the angular resolution and its impact on the determination of the weak mixing angle, we note that the key to reject backgrounds in this study is the kinematic limit $E_e\theta_e^2<2m_e$, where $E_e$, $\theta_e$ and $m_e$ are the outgoing electron energy, angle with respect to the neutrino beam, and mass.
The detector capability to measure energy and angle of recoiled electrons has a direct effect on the $E_e\theta_e^2$ spectrum.
To exemplify that, we show in Fig.~\ref{fig:Specsm} the signal (solid), CCQE background (dashed, magenta) and $\pi^0$ mis-identified background (dashed, cyan) spectra for four assumptions on the angle resolution $\sigma_\theta$, as indicated in the figure for half year of data taking at the on-axis position in the neutrino mode.
We have checked that energy resolution plays a small role in the spectral distortion.
The kinematic limit can be seen as the sharp feature in the perfect resolution histogram around 1~MeV. Clearly, the signal-to-background ratio in the low $E_e\theta_e^2$ bins are worse for worse angular resolution.

\begin{figure}[thbp]
  \centering
  \includegraphics[width=0.32\linewidth]{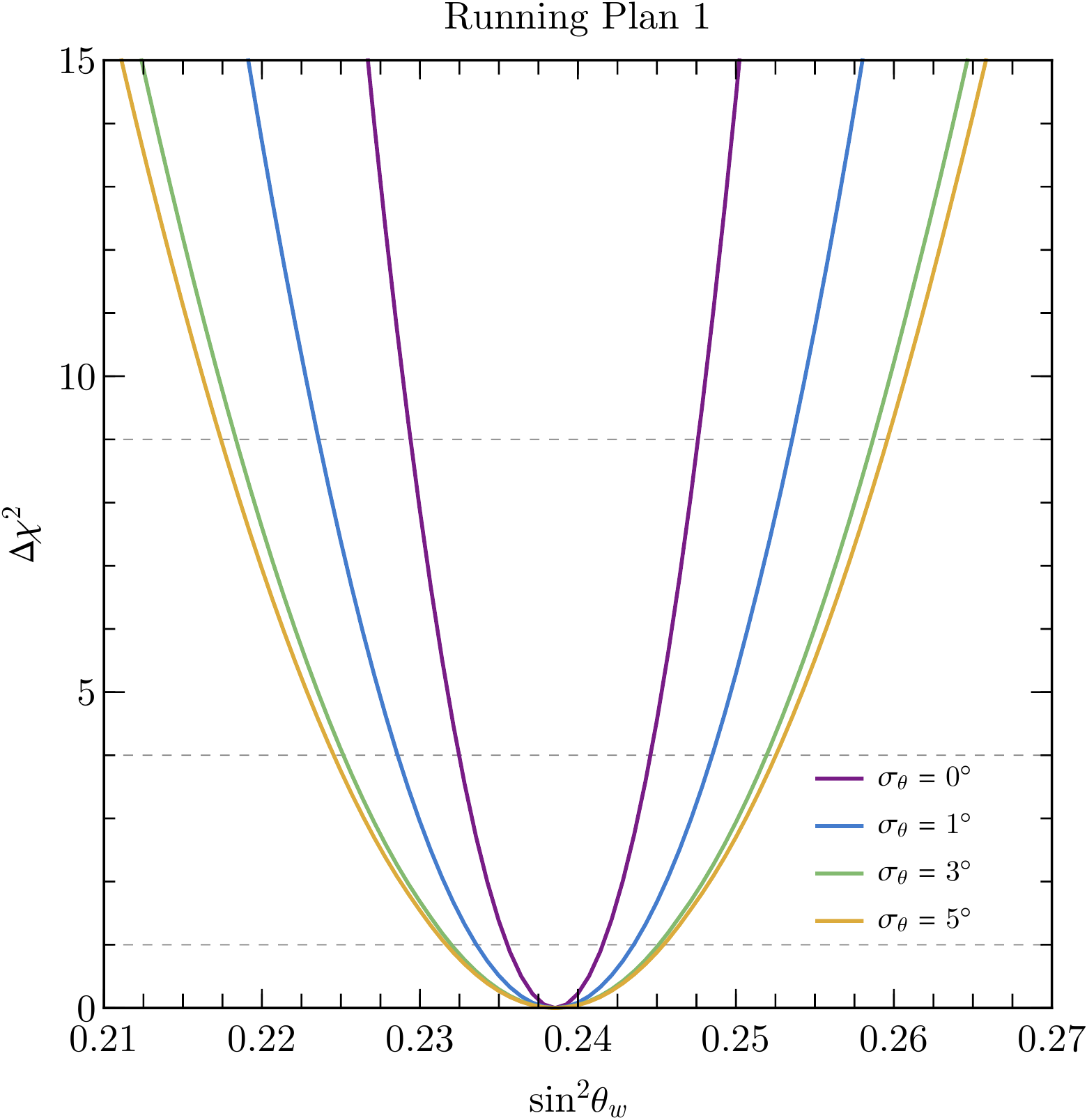}
  \includegraphics[width=0.32\linewidth]{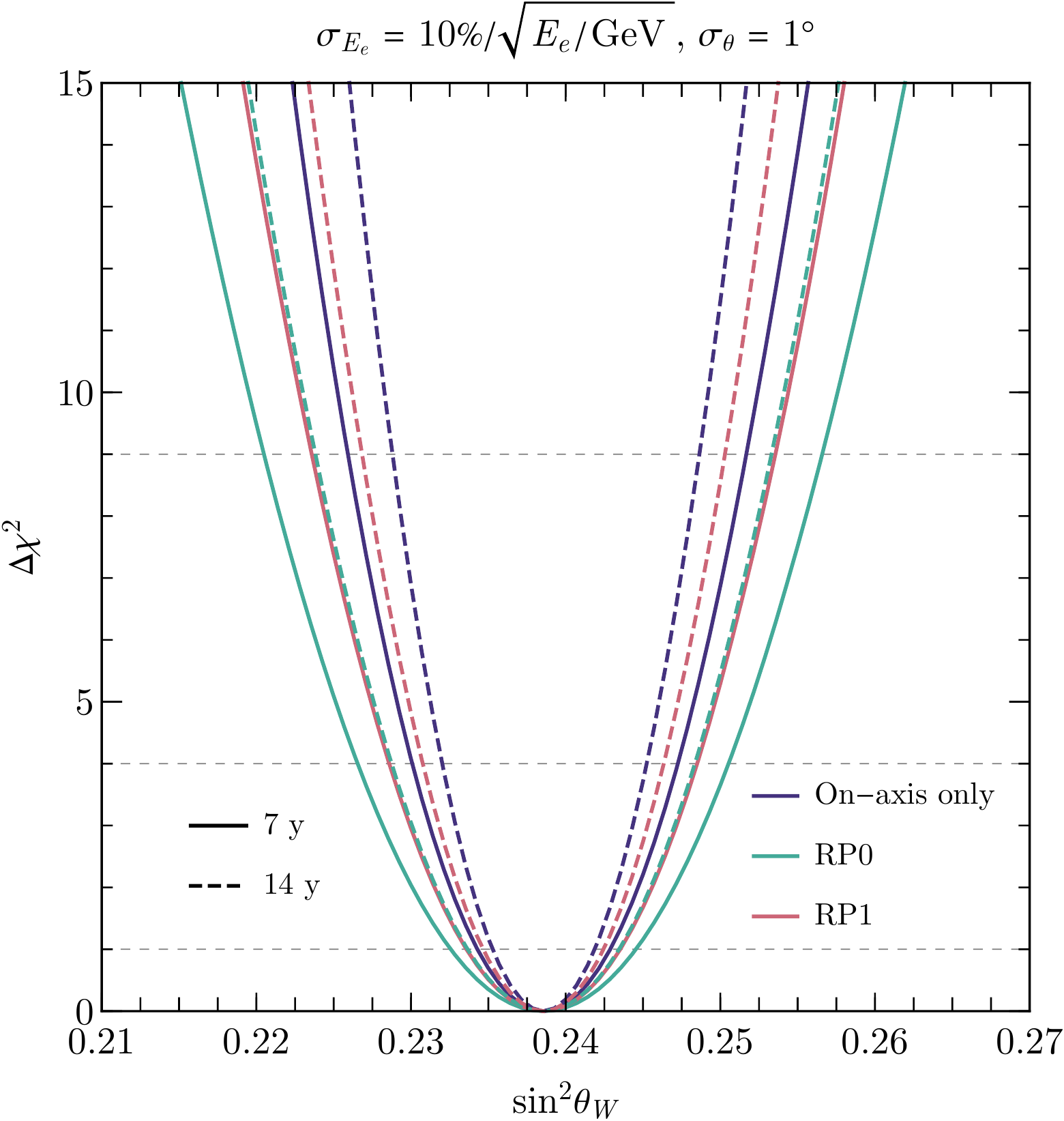}
  \includegraphics[width=0.32\linewidth]{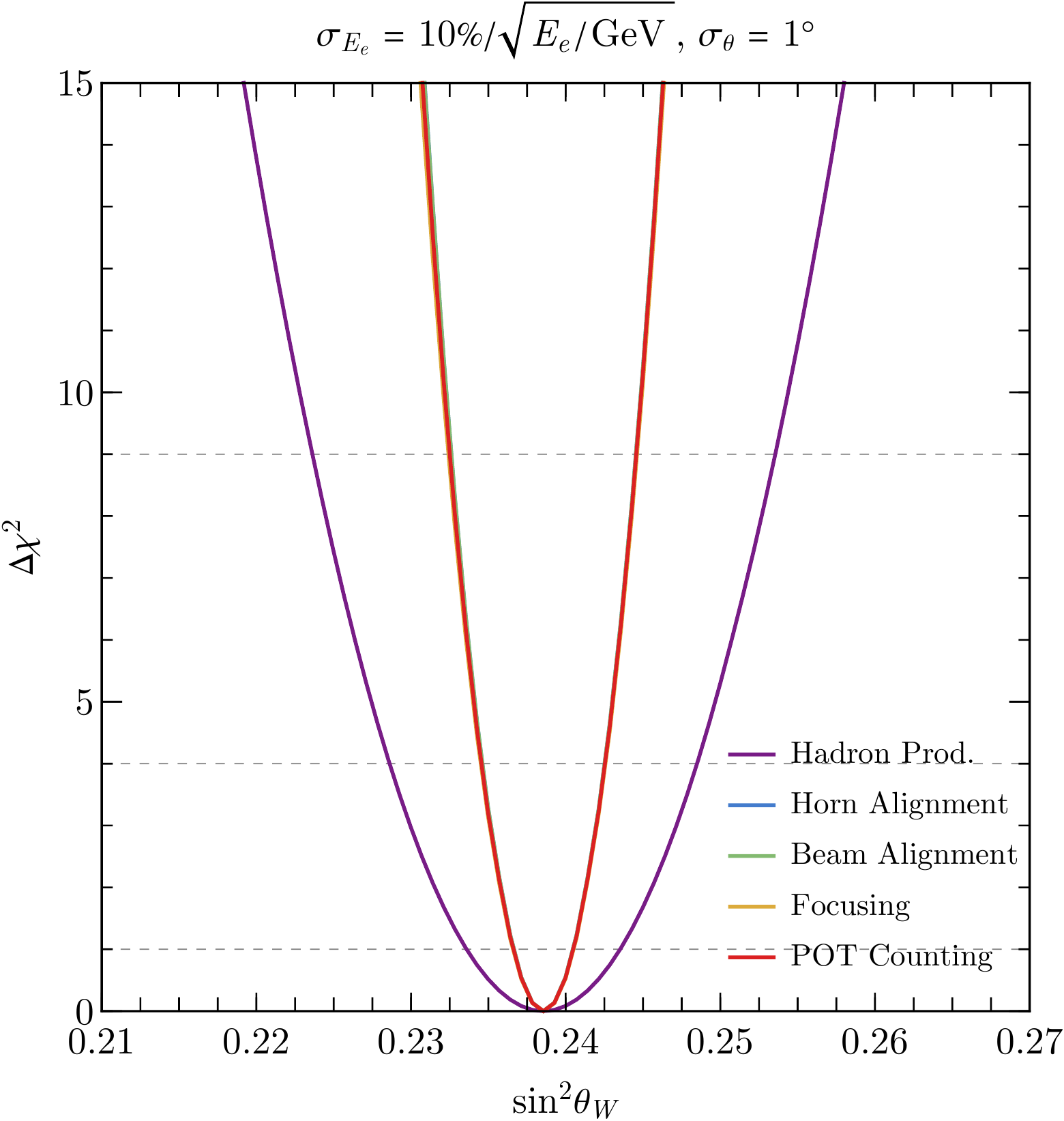}
  \caption{Comparison of the $\Delta\chi^2$ for different values of the resolution of the electron angle $\sigma_\theta$ (left), running plans (center), systematic uncertainties (right).}
  \label{fig:chi2smth}
\end{figure}

To see how the this affects the sensitivity to $\sin^2\theta_W$, we present in the left panel of Fig.~\ref{fig:chi2smth} the $\Delta\chi^2$ as a function of $\sin^2\theta_W$ for several angular resolutions $\sigma_\theta$ as indicated in the figure for the on-axis only, 7 years running. 
The $1\sigma$ uncertainty for $\sin^2\theta_W$ for each assumed angular resolution are $\{1.27\%,2.07\%,2.75\%,2.88\%\}$ for $\sigma_\theta=0^\circ,\,1^\circ,\,3^\circ,\,5^\circ$, respectively.
This indicates that it is important to achieve a good angular resolution in order for DUNE to provide a competitive measurement of the weak mixing angle.

We also show, in the middle panel of  Fig.~\ref{fig:chi2smth} how more statistics or different running plans could affect DUNE's sensitivity to $\sin^2\theta_W$.
The purple, green and magenta lines represent three different running plans: only on-axis; time equally divided in each position (running plan 0 - RP0); and half of the time on-axis and half of the time divided equally in each off-axis position (unning plan 1 - RP1), respectively. 
The solid lines correspond to a total run of 7 years while the dashed lines correspond to a total run of 14 years.
It is clear from the comparison of solid to dashed lines, that the measurement of $\sin^2\theta_W$ is still limited by statistics.
Due to that, spending more time in positions which lead to higher statistics (e.g., on-axis), is beneficial to this measurement.

Finally, to estimate what is the dominant systematic uncertainty that hinders the determination of $\sin^2\theta_W$, we perform the analysis singling out specific systematic uncertainties.
The uncertainties in the neutrino flux come from hadron production, focusing of the beam, the alignment of the beam itself, the number of protons on target (POT), and the horn alignment. 
The results of the analysis for a single uncertainty at a time are shown in the right panel of  Fig.~\ref{fig:chi2smth}.
It is clear that the major uncertainty comes from the hadron production model, while all other sources of uncertainty affect the measurement marginally.

%% file: dune_EM.bbl
\begin{thebibliography}{49}%
\makeatletter
\providecommand \@ifxundefined [1]{%
 \@ifx{#1\undefined}
}%
\providecommand \@ifnum [1]{%
 \ifnum #1\expandafter \@firstoftwo
 \else \expandafter \@secondoftwo
 \fi
}%
\providecommand \@ifx [1]{%
 \ifx #1\expandafter \@firstoftwo
 \else \expandafter \@secondoftwo
 \fi
}%
\providecommand \natexlab [1]{#1}%
\providecommand \enquote  [1]{``#1''}%
\providecommand \bibnamefont  [1]{#1}%
\providecommand \bibfnamefont [1]{#1}%
\providecommand \citenamefont [1]{#1}%
\providecommand \href@noop [0]{\@secondoftwo}%
\providecommand \href [0]{\begingroup \@sanitize@url \@href}%
\providecommand \@href[1]{\@@startlink{#1}\@@href}%
\providecommand \@@href[1]{\endgroup#1\@@endlink}%
\providecommand \@sanitize@url [0]{\catcode `\\12\catcode `\$12\catcode
  `\&12\catcode `\#12\catcode `\^12\catcode `\_12\catcode `\%12\relax}%
\providecommand \@@startlink[1]{}%
\providecommand \@@endlink[0]{}%
\providecommand \url  [0]{\begingroup\@sanitize@url \@url }%
\providecommand \@url [1]{\endgroup\@href {#1}{\urlprefix }}%
\providecommand \urlprefix  [0]{URL }%
\providecommand \Eprint [0]{\href }%
\providecommand \doibase [0]{http://dx.doi.org/}%
\providecommand \selectlanguage [0]{\@gobble}%
\providecommand \bibinfo  [0]{\@secondoftwo}%
\providecommand \bibfield  [0]{\@secondoftwo}%
\providecommand \translation [1]{[#1]}%
\providecommand \BibitemOpen [0]{}%
\providecommand \bibitemStop [0]{}%
\providecommand \bibitemNoStop [0]{.\EOS\space}%
\providecommand \EOS [0]{\spacefactor3000\relax}%
\providecommand \BibitemShut  [1]{\csname bibitem#1\endcsname}%
\let\auto@bib@innerbib\@empty
\bibitem [{\citenamefont {Erler}\ and\ \citenamefont
  {Ramsey-Musolf}(2005)}]{Erler:2004in}%
  \BibitemOpen
  \bibfield  {author} {\bibinfo {author} {\bibfnamefont {J.}~\bibnamefont
  {Erler}}\ and\ \bibinfo {author} {\bibfnamefont {M.~J.}\ \bibnamefont
  {Ramsey-Musolf}},\ }\href {\doibase 10.1103/PhysRevD.72.073003} {\bibfield
  {journal} {\bibinfo  {journal} {Phys. Rev.}\ }\textbf {\bibinfo {volume}
  {D72}},\ \bibinfo {pages} {073003} (\bibinfo {year} {2005})},\ \Eprint
  {http://arxiv.org/abs/hep-ph/0409169} {arXiv:hep-ph/0409169 [hep-ph]}
  \BibitemShut {NoStop}%
\bibitem [{\citenamefont {Erler}\ and\ \citenamefont
  {Ferro-Hern\'{a}ndez}(2018)}]{Erler:2017knj}%
  \BibitemOpen
  \bibfield  {author} {\bibinfo {author} {\bibfnamefont {J.}~\bibnamefont
  {Erler}}\ and\ \bibinfo {author} {\bibfnamefont {R.}~\bibnamefont
  {Ferro-Hern\'{a}ndez}},\ }\href {\doibase 10.1007/JHEP03(2018)196} {\bibfield
   {journal} {\bibinfo  {journal} {JHEP}\ }\textbf {\bibinfo {volume} {03}},\
  \bibinfo {pages} {196} (\bibinfo {year} {2018})},\ \Eprint
  {http://arxiv.org/abs/1712.09146} {arXiv:1712.09146 [hep-ph]} \BibitemShut
  {NoStop}%
\bibitem [{\citenamefont {Tanabashi}\ \emph {et~al.}(2018)\citenamefont
  {Tanabashi} \emph {et~al.}}]{Tanabashi:2018oca}%
  \BibitemOpen
  \bibfield  {author} {\bibinfo {author} {\bibfnamefont {M.}~\bibnamefont
  {Tanabashi}} \emph {et~al.} (\bibinfo {collaboration} {Particle Data
  Group}),\ }\href {\doibase 10.1103/PhysRevD.98.030001} {\bibfield  {journal}
  {\bibinfo  {journal} {Phys. Rev.}\ }\textbf {\bibinfo {volume} {D98}},\
  \bibinfo {pages} {030001} (\bibinfo {year} {2018})}\BibitemShut {NoStop}%
\bibitem [{\citenamefont {Zeller}\ \emph {et~al.}(2002)\citenamefont {Zeller}
  \emph {et~al.}}]{Zeller:2001hh}%
  \BibitemOpen
  \bibfield  {author} {\bibinfo {author} {\bibfnamefont {G.~P.}\ \bibnamefont
  {Zeller}} \emph {et~al.} (\bibinfo {collaboration} {NuTeV}),\ }\href
  {\doibase 10.1103/PhysRevLett.88.091802, 10.1103/PhysRevLett.90.239902}
  {\bibfield  {journal} {\bibinfo  {journal} {Phys. Rev. Lett.}\ }\textbf
  {\bibinfo {volume} {88}},\ \bibinfo {pages} {091802} (\bibinfo {year}
  {2002})},\ \bibinfo {note} {[Erratum: Phys. Rev. Lett.90,239902(2003)]},\
  \Eprint {http://arxiv.org/abs/hep-ex/0110059} {arXiv:hep-ex/0110059 [hep-ex]}
  \BibitemShut {NoStop}%
\bibitem [{\citenamefont {Schael}\ \emph {et~al.}(2006)\citenamefont {Schael}
  \emph {et~al.}}]{ALEPH:2005ab}%
  \BibitemOpen
  \bibfield  {author} {\bibinfo {author} {\bibfnamefont {S.}~\bibnamefont
  {Schael}} \emph {et~al.} (\bibinfo {collaboration} {ALEPH, DELPHI, L3, OPAL,
  SLD, LEP Electroweak Working Group, SLD Electroweak Group, SLD Heavy Flavour
  Group}),\ }\href {\doibase 10.1016/j.physrep.2005.12.006} {\bibfield
  {journal} {\bibinfo  {journal} {Phys. Rept.}\ }\textbf {\bibinfo {volume}
  {427}},\ \bibinfo {pages} {257} (\bibinfo {year} {2006})},\ \Eprint
  {http://arxiv.org/abs/hep-ex/0509008} {arXiv:hep-ex/0509008 [hep-ex]}
  \BibitemShut {NoStop}%
\bibitem [{\citenamefont {Pumplin}\ \emph {et~al.}(2002)\citenamefont
  {Pumplin}, \citenamefont {Stump}, \citenamefont {Huston}, \citenamefont
  {Lai}, \citenamefont {Nadolsky},\ and\ \citenamefont
  {Tung}}]{Pumplin:2002vw}%
  \BibitemOpen
  \bibfield  {author} {\bibinfo {author} {\bibfnamefont {J.}~\bibnamefont
  {Pumplin}}, \bibinfo {author} {\bibfnamefont {D.~R.}\ \bibnamefont {Stump}},
  \bibinfo {author} {\bibfnamefont {J.}~\bibnamefont {Huston}}, \bibinfo
  {author} {\bibfnamefont {H.~L.}\ \bibnamefont {Lai}}, \bibinfo {author}
  {\bibfnamefont {P.~M.}\ \bibnamefont {Nadolsky}}, \ and\ \bibinfo {author}
  {\bibfnamefont {W.~K.}\ \bibnamefont {Tung}},\ }\href {\doibase
  10.1088/1126-6708/2002/07/012} {\bibfield  {journal} {\bibinfo  {journal}
  {JHEP}\ }\textbf {\bibinfo {volume} {07}},\ \bibinfo {pages} {012} (\bibinfo
  {year} {2002})},\ \Eprint {http://arxiv.org/abs/hep-ph/0201195}
  {arXiv:hep-ph/0201195 [hep-ph]} \BibitemShut {NoStop}%
\bibitem [{\citenamefont {Kretzer}\ \emph {et~al.}(2004)\citenamefont
  {Kretzer}, \citenamefont {Olness}, \citenamefont {Pumplin}, \citenamefont
  {Stump}, \citenamefont {Tung},\ and\ \citenamefont {Reno}}]{Kretzer:2003wy}%
  \BibitemOpen
  \bibfield  {author} {\bibinfo {author} {\bibfnamefont {S.}~\bibnamefont
  {Kretzer}}, \bibinfo {author} {\bibfnamefont {F.}~\bibnamefont {Olness}},
  \bibinfo {author} {\bibfnamefont {J.}~\bibnamefont {Pumplin}}, \bibinfo
  {author} {\bibfnamefont {D.}~\bibnamefont {Stump}}, \bibinfo {author}
  {\bibfnamefont {W.-K.}\ \bibnamefont {Tung}}, \ and\ \bibinfo {author}
  {\bibfnamefont {M.~H.}\ \bibnamefont {Reno}},\ }\href {\doibase
  10.1103/PhysRevLett.93.041802} {\bibfield  {journal} {\bibinfo  {journal}
  {Phys. Rev. Lett.}\ }\textbf {\bibinfo {volume} {93}},\ \bibinfo {pages}
  {041802} (\bibinfo {year} {2004})},\ \Eprint
  {http://arxiv.org/abs/hep-ph/0312322} {arXiv:hep-ph/0312322 [hep-ph]}
  \BibitemShut {NoStop}%
\bibitem [{\citenamefont {Sather}(1992)}]{Sather:1991je}%
  \BibitemOpen
  \bibfield  {author} {\bibinfo {author} {\bibfnamefont {E.}~\bibnamefont
  {Sather}},\ }\href {\doibase 10.1016/0370-2693(92)92011-5} {\bibfield
  {journal} {\bibinfo  {journal} {Phys. Lett.}\ }\textbf {\bibinfo {volume}
  {B274}},\ \bibinfo {pages} {433} (\bibinfo {year} {1992})}\BibitemShut
  {NoStop}%
\bibitem [{\citenamefont {Rodionov}\ \emph {et~al.}(1994)\citenamefont
  {Rodionov}, \citenamefont {Thomas},\ and\ \citenamefont
  {Londergan}}]{Rodionov:1994cg}%
  \BibitemOpen
  \bibfield  {author} {\bibinfo {author} {\bibfnamefont {E.~N.}\ \bibnamefont
  {Rodionov}}, \bibinfo {author} {\bibfnamefont {A.~W.}\ \bibnamefont
  {Thomas}}, \ and\ \bibinfo {author} {\bibfnamefont {J.~T.}\ \bibnamefont
  {Londergan}},\ }\href {\doibase 10.1142/S0217732394001659} {\bibfield
  {journal} {\bibinfo  {journal} {Mod. Phys. Lett.}\ }\textbf {\bibinfo
  {volume} {A9}},\ \bibinfo {pages} {1799} (\bibinfo {year}
  {1994})}\BibitemShut {NoStop}%
\bibitem [{\citenamefont {Martin}\ \emph {et~al.}(2004)\citenamefont {Martin},
  \citenamefont {Roberts}, \citenamefont {Stirling},\ and\ \citenamefont
  {Thorne}}]{Martin:2003sk}%
  \BibitemOpen
  \bibfield  {author} {\bibinfo {author} {\bibfnamefont {A.~D.}\ \bibnamefont
  {Martin}}, \bibinfo {author} {\bibfnamefont {R.~G.}\ \bibnamefont {Roberts}},
  \bibinfo {author} {\bibfnamefont {W.~J.}\ \bibnamefont {Stirling}}, \ and\
  \bibinfo {author} {\bibfnamefont {R.~S.}\ \bibnamefont {Thorne}},\ }\href
  {\doibase 10.1140/epjc/s2004-01825-2} {\bibfield  {journal} {\bibinfo
  {journal} {Eur. Phys. J.}\ }\textbf {\bibinfo {volume} {C35}},\ \bibinfo
  {pages} {325} (\bibinfo {year} {2004})},\ \Eprint
  {http://arxiv.org/abs/hep-ph/0308087} {arXiv:hep-ph/0308087 [hep-ph]}
  \BibitemShut {NoStop}%
\bibitem [{\citenamefont {Londergan}\ and\ \citenamefont
  {Thomas}(2003)}]{Londergan:2003ij}%
  \BibitemOpen
  \bibfield  {author} {\bibinfo {author} {\bibfnamefont {J.~T.}\ \bibnamefont
  {Londergan}}\ and\ \bibinfo {author} {\bibfnamefont {A.~W.}\ \bibnamefont
  {Thomas}},\ }\href {\doibase 10.1103/PhysRevD.67.111901} {\bibfield
  {journal} {\bibinfo  {journal} {Phys. Rev.}\ }\textbf {\bibinfo {volume}
  {D67}},\ \bibinfo {pages} {111901} (\bibinfo {year} {2003})},\ \Eprint
  {http://arxiv.org/abs/hep-ph/0303155} {arXiv:hep-ph/0303155 [hep-ph]}
  \BibitemShut {NoStop}%
\bibitem [{\citenamefont {Bentz}\ \emph {et~al.}(2010)\citenamefont {Bentz},
  \citenamefont {Cloet}, \citenamefont {Londergan},\ and\ \citenamefont
  {Thomas}}]{Bentz:2009yy}%
  \BibitemOpen
  \bibfield  {author} {\bibinfo {author} {\bibfnamefont {W.}~\bibnamefont
  {Bentz}}, \bibinfo {author} {\bibfnamefont {I.~C.}\ \bibnamefont {Cloet}},
  \bibinfo {author} {\bibfnamefont {J.~T.}\ \bibnamefont {Londergan}}, \ and\
  \bibinfo {author} {\bibfnamefont {A.~W.}\ \bibnamefont {Thomas}},\ }\href
  {\doibase 10.1016/j.physletb.2010.09.001} {\bibfield  {journal} {\bibinfo
  {journal} {Phys. Lett.}\ }\textbf {\bibinfo {volume} {B693}},\ \bibinfo
  {pages} {462} (\bibinfo {year} {2010})},\ \Eprint
  {http://arxiv.org/abs/0908.3198} {arXiv:0908.3198 [nucl-th]} \BibitemShut
  {NoStop}%
\bibitem [{\citenamefont {Gluck}\ \emph {et~al.}(2005)\citenamefont {Gluck},
  \citenamefont {Jimenez-Delgado},\ and\ \citenamefont {Reya}}]{Gluck:2005xh}%
  \BibitemOpen
  \bibfield  {author} {\bibinfo {author} {\bibfnamefont {M.}~\bibnamefont
  {Gluck}}, \bibinfo {author} {\bibfnamefont {P.}~\bibnamefont
  {Jimenez-Delgado}}, \ and\ \bibinfo {author} {\bibfnamefont {E.}~\bibnamefont
  {Reya}},\ }\href {\doibase 10.1103/PhysRevLett.95.022002} {\bibfield
  {journal} {\bibinfo  {journal} {Phys. Rev. Lett.}\ }\textbf {\bibinfo
  {volume} {95}},\ \bibinfo {pages} {022002} (\bibinfo {year} {2005})},\
  \Eprint {http://arxiv.org/abs/hep-ph/0503103} {arXiv:hep-ph/0503103 [hep-ph]}
  \BibitemShut {NoStop}%
\bibitem [{\citenamefont {Kumano}(2002)}]{Kumano:2002ra}%
  \BibitemOpen
  \bibfield  {author} {\bibinfo {author} {\bibfnamefont {S.}~\bibnamefont
  {Kumano}},\ }\href {\doibase 10.1103/PhysRevD.66.111301} {\bibfield
  {journal} {\bibinfo  {journal} {Phys. Rev.}\ }\textbf {\bibinfo {volume}
  {D66}},\ \bibinfo {pages} {111301} (\bibinfo {year} {2002})},\ \Eprint
  {http://arxiv.org/abs/hep-ph/0209200} {arXiv:hep-ph/0209200 [hep-ph]}
  \BibitemShut {NoStop}%
\bibitem [{\citenamefont {Kulagin}(2003)}]{Kulagin:2003wz}%
  \BibitemOpen
  \bibfield  {author} {\bibinfo {author} {\bibfnamefont {S.~A.}\ \bibnamefont
  {Kulagin}},\ }\href {\doibase 10.1103/PhysRevD.67.091301} {\bibfield
  {journal} {\bibinfo  {journal} {Phys. Rev.}\ }\textbf {\bibinfo {volume}
  {D67}},\ \bibinfo {pages} {091301} (\bibinfo {year} {2003})},\ \Eprint
  {http://arxiv.org/abs/hep-ph/0301045} {arXiv:hep-ph/0301045 [hep-ph]}
  \BibitemShut {NoStop}%
\bibitem [{\citenamefont {Brodsky}\ \emph {et~al.}(2004)\citenamefont
  {Brodsky}, \citenamefont {Schmidt},\ and\ \citenamefont
  {Yang}}]{Brodsky:2004qa}%
  \BibitemOpen
  \bibfield  {author} {\bibinfo {author} {\bibfnamefont {S.~J.}\ \bibnamefont
  {Brodsky}}, \bibinfo {author} {\bibfnamefont {I.}~\bibnamefont {Schmidt}}, \
  and\ \bibinfo {author} {\bibfnamefont {J.-J.}\ \bibnamefont {Yang}},\ }\href
  {\doibase 10.1103/PhysRevD.70.116003} {\bibfield  {journal} {\bibinfo
  {journal} {Phys. Rev.}\ }\textbf {\bibinfo {volume} {D70}},\ \bibinfo {pages}
  {116003} (\bibinfo {year} {2004})},\ \Eprint
  {http://arxiv.org/abs/hep-ph/0409279} {arXiv:hep-ph/0409279 [hep-ph]}
  \BibitemShut {NoStop}%
\bibitem [{\citenamefont {Hirai}\ \emph {et~al.}(2005)\citenamefont {Hirai},
  \citenamefont {Kumano},\ and\ \citenamefont {Nagai}}]{Hirai:2004ba}%
  \BibitemOpen
  \bibfield  {author} {\bibinfo {author} {\bibfnamefont {M.}~\bibnamefont
  {Hirai}}, \bibinfo {author} {\bibfnamefont {S.}~\bibnamefont {Kumano}}, \
  and\ \bibinfo {author} {\bibfnamefont {T.~H.}\ \bibnamefont {Nagai}},\ }\href
  {\doibase 10.1103/PhysRevD.71.113007} {\bibfield  {journal} {\bibinfo
  {journal} {Phys. Rev.}\ }\textbf {\bibinfo {volume} {D71}},\ \bibinfo {pages}
  {113007} (\bibinfo {year} {2005})},\ \Eprint
  {http://arxiv.org/abs/hep-ph/0412284} {arXiv:hep-ph/0412284 [hep-ph]}
  \BibitemShut {NoStop}%
\bibitem [{\citenamefont {Miller}\ and\ \citenamefont
  {Thomas}(2005)}]{Miller:2002xh}%
  \BibitemOpen
  \bibfield  {author} {\bibinfo {author} {\bibfnamefont {G.~A.}\ \bibnamefont
  {Miller}}\ and\ \bibinfo {author} {\bibfnamefont {A.~W.}\ \bibnamefont
  {Thomas}},\ }\href {\doibase 10.1142/S0217751X0502121X} {\bibfield  {journal}
  {\bibinfo  {journal} {Int. J. Mod. Phys.}\ }\textbf {\bibinfo {volume}
  {A20}},\ \bibinfo {pages} {95} (\bibinfo {year} {2005})},\ \Eprint
  {http://arxiv.org/abs/hep-ex/0204007} {arXiv:hep-ex/0204007 [hep-ex]}
  \BibitemShut {NoStop}%
\bibitem [{\citenamefont {Cloet}\ \emph {et~al.}(2009)\citenamefont {Cloet},
  \citenamefont {Bentz},\ and\ \citenamefont {Thomas}}]{Cloet:2009qs}%
  \BibitemOpen
  \bibfield  {author} {\bibinfo {author} {\bibfnamefont {I.~C.}\ \bibnamefont
  {Cloet}}, \bibinfo {author} {\bibfnamefont {W.}~\bibnamefont {Bentz}}, \ and\
  \bibinfo {author} {\bibfnamefont {A.~W.}\ \bibnamefont {Thomas}},\ }\href
  {\doibase 10.1103/PhysRevLett.102.252301} {\bibfield  {journal} {\bibinfo
  {journal} {Phys. Rev. Lett.}\ }\textbf {\bibinfo {volume} {102}},\ \bibinfo
  {pages} {252301} (\bibinfo {year} {2009})},\ \Eprint
  {http://arxiv.org/abs/0901.3559} {arXiv:0901.3559 [nucl-th]} \BibitemShut
  {NoStop}%
\bibitem [{\citenamefont {Diener}\ \emph {et~al.}(2004)\citenamefont {Diener},
  \citenamefont {Dittmaier},\ and\ \citenamefont {Hollik}}]{Diener:2003ss}%
  \BibitemOpen
  \bibfield  {author} {\bibinfo {author} {\bibfnamefont {K.~P.~O.}\
  \bibnamefont {Diener}}, \bibinfo {author} {\bibfnamefont {S.}~\bibnamefont
  {Dittmaier}}, \ and\ \bibinfo {author} {\bibfnamefont {W.}~\bibnamefont
  {Hollik}},\ }\href {\doibase 10.1103/PhysRevD.69.073005} {\bibfield
  {journal} {\bibinfo  {journal} {Phys. Rev.}\ }\textbf {\bibinfo {volume}
  {D69}},\ \bibinfo {pages} {073005} (\bibinfo {year} {2004})},\ \Eprint
  {http://arxiv.org/abs/hep-ph/0310364} {arXiv:hep-ph/0310364 [hep-ph]}
  \BibitemShut {NoStop}%
\bibitem [{\citenamefont {Arbuzov}\ \emph {et~al.}(2005)\citenamefont
  {Arbuzov}, \citenamefont {Bardin},\ and\ \citenamefont
  {Kalinovskaya}}]{Arbuzov:2004zr}%
  \BibitemOpen
  \bibfield  {author} {\bibinfo {author} {\bibfnamefont {A.~B.}\ \bibnamefont
  {Arbuzov}}, \bibinfo {author} {\bibfnamefont {D.~{\relax Yu}.}\ \bibnamefont
  {Bardin}}, \ and\ \bibinfo {author} {\bibfnamefont {L.~V.}\ \bibnamefont
  {Kalinovskaya}},\ }\href {\doibase 10.1088/1126-6708/2005/06/078} {\bibfield
  {journal} {\bibinfo  {journal} {JHEP}\ }\textbf {\bibinfo {volume} {06}},\
  \bibinfo {pages} {078} (\bibinfo {year} {2005})},\ \Eprint
  {http://arxiv.org/abs/hep-ph/0407203} {arXiv:hep-ph/0407203 [hep-ph]}
  \BibitemShut {NoStop}%
\bibitem [{\citenamefont {Park}\ \emph {et~al.}(2009)\citenamefont {Park},
  \citenamefont {Baur},\ and\ \citenamefont {Wackeroth}}]{Park:2009ft}%
  \BibitemOpen
  \bibfield  {author} {\bibinfo {author} {\bibfnamefont {K.}~\bibnamefont
  {Park}}, \bibinfo {author} {\bibfnamefont {U.}~\bibnamefont {Baur}}, \ and\
  \bibinfo {author} {\bibfnamefont {D.}~\bibnamefont {Wackeroth}},\ }in\
  \href@noop {} {\emph {\bibinfo {booktitle} {{Particles and fields.
  Proceedings, Meeting of the Division of the American Physical Society, DPF
  2009, Detroit, USA, July 26-31, 2009}}}}\ (\bibinfo {year} {2009})\ \Eprint
  {http://arxiv.org/abs/0910.5013} {arXiv:0910.5013 [hep-ph]} \BibitemShut
  {NoStop}%
\bibitem [{\citenamefont {Diener}\ \emph {et~al.}(2005)\citenamefont {Diener},
  \citenamefont {Dittmaier},\ and\ \citenamefont {Hollik}}]{Diener:2005me}%
  \BibitemOpen
  \bibfield  {author} {\bibinfo {author} {\bibfnamefont {K.~P.~O.}\
  \bibnamefont {Diener}}, \bibinfo {author} {\bibfnamefont {S.}~\bibnamefont
  {Dittmaier}}, \ and\ \bibinfo {author} {\bibfnamefont {W.}~\bibnamefont
  {Hollik}},\ }\href {\doibase 10.1103/PhysRevD.72.093002} {\bibfield
  {journal} {\bibinfo  {journal} {Phys. Rev.}\ }\textbf {\bibinfo {volume}
  {D72}},\ \bibinfo {pages} {093002} (\bibinfo {year} {2005})},\ \Eprint
  {http://arxiv.org/abs/hep-ph/0509084} {arXiv:hep-ph/0509084 [hep-ph]}
  \BibitemShut {NoStop}%
\bibitem [{\citenamefont {Dobrescu}\ and\ \citenamefont
  {Ellis}(2004)}]{Dobrescu:2003ta}%
  \BibitemOpen
  \bibfield  {author} {\bibinfo {author} {\bibfnamefont {B.~A.}\ \bibnamefont
  {Dobrescu}}\ and\ \bibinfo {author} {\bibfnamefont {R.~K.}\ \bibnamefont
  {Ellis}},\ }\href {\doibase 10.1103/PhysRevD.69.114014} {\bibfield  {journal}
  {\bibinfo  {journal} {Phys. Rev.}\ }\textbf {\bibinfo {volume} {D69}},\
  \bibinfo {pages} {114014} (\bibinfo {year} {2004})},\ \Eprint
  {http://arxiv.org/abs/hep-ph/0310154} {arXiv:hep-ph/0310154 [hep-ph]}
  \BibitemShut {NoStop}%
\bibitem [{\citenamefont {Acciarri}\ \emph {et~al.}(2015)\citenamefont
  {Acciarri} \emph {et~al.}}]{Acciarri:2015uup}%
  \BibitemOpen
  \bibfield  {author} {\bibinfo {author} {\bibfnamefont {R.}~\bibnamefont
  {Acciarri}} \emph {et~al.} (\bibinfo {collaboration} {DUNE}),\ }\href@noop {}
  {\  (\bibinfo {year} {2015})},\ \Eprint {http://arxiv.org/abs/1512.06148}
  {arXiv:1512.06148 [physics.ins-det]} \BibitemShut {NoStop}%
\bibitem [{\citenamefont {Abe}\ \emph {et~al.}(2011)\citenamefont {Abe} \emph
  {et~al.}}]{Abe:2011ts}%
  \BibitemOpen
  \bibfield  {author} {\bibinfo {author} {\bibfnamefont {K.}~\bibnamefont
  {Abe}} \emph {et~al.},\ }\href@noop {} {\  (\bibinfo {year} {2011})},\
  \Eprint {http://arxiv.org/abs/1109.3262} {arXiv:1109.3262 [hep-ex]}
  \BibitemShut {NoStop}%
\bibitem [{\citenamefont {Alvarez-Ruso}\ \emph {et~al.}(2018)\citenamefont
  {Alvarez-Ruso} \emph {et~al.}}]{Alvarez-Ruso:2017oui}%
  \BibitemOpen
  \bibfield  {author} {\bibinfo {author} {\bibfnamefont {L.}~\bibnamefont
  {Alvarez-Ruso}} \emph {et~al.},\ }\href {\doibase 10.1016/j.ppnp.2018.01.006}
  {\bibfield  {journal} {\bibinfo  {journal} {Prog. Part. Nucl. Phys.}\
  }\textbf {\bibinfo {volume} {100}},\ \bibinfo {pages} {1} (\bibinfo {year}
  {2018})},\ \Eprint {http://arxiv.org/abs/1706.03621} {arXiv:1706.03621
  [hep-ph]} \BibitemShut {NoStop}%
\bibitem [{\citenamefont {Conrad}\ \emph {et~al.}(2005)\citenamefont {Conrad},
  \citenamefont {Link},\ and\ \citenamefont {Shaevitz}}]{Conrad:2004gw}%
  \BibitemOpen
  \bibfield  {author} {\bibinfo {author} {\bibfnamefont {J.~M.}\ \bibnamefont
  {Conrad}}, \bibinfo {author} {\bibfnamefont {J.~M.}\ \bibnamefont {Link}}, \
  and\ \bibinfo {author} {\bibfnamefont {M.~H.}\ \bibnamefont {Shaevitz}},\
  }\href {\doibase 10.1103/PhysRevD.71.073013} {\bibfield  {journal} {\bibinfo
  {journal} {Phys. Rev.}\ }\textbf {\bibinfo {volume} {D71}},\ \bibinfo {pages}
  {073013} (\bibinfo {year} {2005})},\ \Eprint
  {http://arxiv.org/abs/hep-ex/0403048} {arXiv:hep-ex/0403048 [hep-ex]}
  \BibitemShut {NoStop}%
\bibitem [{\citenamefont {de~Gouv\^ea}\ and\ \citenamefont
  {Jenkins}(2006)}]{deGouvea:2006hfo}%
  \BibitemOpen
  \bibfield  {author} {\bibinfo {author} {\bibfnamefont {A.}~\bibnamefont
  {de~Gouv\^ea}}\ and\ \bibinfo {author} {\bibfnamefont {J.}~\bibnamefont
  {Jenkins}},\ }\href {\doibase 10.1103/PhysRevD.74.033004} {\bibfield
  {journal} {\bibinfo  {journal} {Phys. Rev.}\ }\textbf {\bibinfo {volume}
  {D74}},\ \bibinfo {pages} {033004} (\bibinfo {year} {2006})},\ \Eprint
  {http://arxiv.org/abs/hep-ph/0603036} {arXiv:hep-ph/0603036 [hep-ph]}
  \BibitemShut {NoStop}%
\bibitem [{\citenamefont {Agarwalla}\ and\ \citenamefont
  {Huber}(2011)}]{Agarwalla:2010ty}%
  \BibitemOpen
  \bibfield  {author} {\bibinfo {author} {\bibfnamefont {S.~K.}\ \bibnamefont
  {Agarwalla}}\ and\ \bibinfo {author} {\bibfnamefont {P.}~\bibnamefont
  {Huber}},\ }\href {\doibase 10.1007/JHEP08(2011)059} {\bibfield  {journal}
  {\bibinfo  {journal} {JHEP}\ }\textbf {\bibinfo {volume} {08}},\ \bibinfo
  {pages} {059} (\bibinfo {year} {2011})},\ \Eprint
  {http://arxiv.org/abs/1005.1254} {arXiv:1005.1254 [hep-ph]} \BibitemShut
  {NoStop}%
\bibitem [{\citenamefont {Conrad}\ \emph {et~al.}(2014)\citenamefont {Conrad},
  \citenamefont {Shaevitz}, \citenamefont {Shimizu}, \citenamefont {Spitz},
  \citenamefont {Toups},\ and\ \citenamefont {Winslow}}]{Conrad:2013sqa}%
  \BibitemOpen
  \bibfield  {author} {\bibinfo {author} {\bibfnamefont {J.~M.}\ \bibnamefont
  {Conrad}}, \bibinfo {author} {\bibfnamefont {M.~H.}\ \bibnamefont
  {Shaevitz}}, \bibinfo {author} {\bibfnamefont {I.}~\bibnamefont {Shimizu}},
  \bibinfo {author} {\bibfnamefont {J.}~\bibnamefont {Spitz}}, \bibinfo
  {author} {\bibfnamefont {M.}~\bibnamefont {Toups}}, \ and\ \bibinfo {author}
  {\bibfnamefont {L.}~\bibnamefont {Winslow}},\ }\href {\doibase
  10.1103/PhysRevD.89.072010} {\bibfield  {journal} {\bibinfo  {journal} {Phys.
  Rev.}\ }\textbf {\bibinfo {volume} {D89}},\ \bibinfo {pages} {072010}
  (\bibinfo {year} {2014})},\ \Eprint {http://arxiv.org/abs/1307.5081}
  {arXiv:1307.5081 [hep-ex]} \BibitemShut {NoStop}%
\bibitem [{\citenamefont {Adelmann}\ \emph {et~al.}(2014)\citenamefont
  {Adelmann}, \citenamefont {Alonso}, \citenamefont {Barletta}, \citenamefont
  {Conrad}, \citenamefont {Shaevitz}, \citenamefont {Spitz}, \citenamefont
  {Toups},\ and\ \citenamefont {Winslow}}]{Adelmann:2013isa}%
  \BibitemOpen
  \bibfield  {author} {\bibinfo {author} {\bibfnamefont {A.}~\bibnamefont
  {Adelmann}}, \bibinfo {author} {\bibfnamefont {J.}~\bibnamefont {Alonso}},
  \bibinfo {author} {\bibfnamefont {W.~A.}\ \bibnamefont {Barletta}}, \bibinfo
  {author} {\bibfnamefont {J.~M.}\ \bibnamefont {Conrad}}, \bibinfo {author}
  {\bibfnamefont {M.~H.}\ \bibnamefont {Shaevitz}}, \bibinfo {author}
  {\bibfnamefont {J.}~\bibnamefont {Spitz}}, \bibinfo {author} {\bibfnamefont
  {M.}~\bibnamefont {Toups}}, \ and\ \bibinfo {author} {\bibfnamefont {L.~A.}\
  \bibnamefont {Winslow}},\ }\href {\doibase 10.1155/2014/347097} {\bibfield
  {journal} {\bibinfo  {journal} {Adv. High Energy Phys.}\ }\textbf {\bibinfo
  {volume} {2014}},\ \bibinfo {pages} {347097} (\bibinfo {year} {2014})},\
  \Eprint {http://arxiv.org/abs/1307.6465} {arXiv:1307.6465 [physics.acc-ph]}
  \BibitemShut {NoStop}%
\bibitem [{\citenamefont {Abe}\ \emph {et~al.}(2013)\citenamefont {Abe} \emph
  {et~al.}}]{Abe:2012av}%
  \BibitemOpen
  \bibfield  {author} {\bibinfo {author} {\bibfnamefont {K.}~\bibnamefont
  {Abe}} \emph {et~al.} (\bibinfo {collaboration} {T2K}),\ }\href {\doibase
  10.1103/PhysRevD.87.012001, 10.1103/PhysRevD.87.019902} {\bibfield  {journal}
  {\bibinfo  {journal} {Phys. Rev.}\ }\textbf {\bibinfo {volume} {D87}},\
  \bibinfo {pages} {012001} (\bibinfo {year} {2013})},\ \bibinfo {note}
  {[Addendum: Phys. Rev.D87,no.1,019902(2013)]},\ \Eprint
  {http://arxiv.org/abs/1211.0469} {arXiv:1211.0469 [hep-ex]} \BibitemShut
  {NoStop}%
\bibitem [{\citenamefont {Aliaga}\ \emph {et~al.}(2016)\citenamefont {Aliaga}
  \emph {et~al.}}]{Aliaga:2016oaz}%
  \BibitemOpen
  \bibfield  {author} {\bibinfo {author} {\bibfnamefont {L.}~\bibnamefont
  {Aliaga}} \emph {et~al.} (\bibinfo {collaboration} {MINERvA}),\ }\href
  {\doibase 10.1103/PhysRevD.94.092005, 10.1103/PhysRevD.95.039903} {\bibfield
  {journal} {\bibinfo  {journal} {Phys. Rev.}\ }\textbf {\bibinfo {volume}
  {D94}},\ \bibinfo {pages} {092005} (\bibinfo {year} {2016})},\ \bibinfo
  {note} {[Addendum: Phys. Rev.D95,no.3,039903(2017)]},\ \Eprint
  {http://arxiv.org/abs/1607.00704} {arXiv:1607.00704 [hep-ex]} \BibitemShut
  {NoStop}%
\bibitem [{\citenamefont {Marshall}\ \emph {et~al.}(2019)\citenamefont
  {Marshall}, \citenamefont {McFarland},\ and\ \citenamefont
  {Wilkinson}}]{Marshall:2019vdy}%
  \BibitemOpen
  \bibfield  {author} {\bibinfo {author} {\bibfnamefont {C.~M.}\ \bibnamefont
  {Marshall}}, \bibinfo {author} {\bibfnamefont {K.~S.}\ \bibnamefont
  {McFarland}}, \ and\ \bibinfo {author} {\bibfnamefont {C.}~\bibnamefont
  {Wilkinson}},\ }\href@noop {} {\  (\bibinfo {year} {2019})},\ \Eprint
  {http://arxiv.org/abs/1910.10996} {arXiv:1910.10996 [hep-ex]} \BibitemShut
  {NoStop}%
\bibitem [{\citenamefont {Pickering}(2019)}]{pickering-prism}%
  \BibitemOpen
  \bibfield  {author} {\bibinfo {author} {\bibfnamefont {L.}~\bibnamefont
  {Pickering}},\ }\href@noop {} {\enquote {\bibinfo {title} {{DUNE-PRISM
  analysis update}},}\ } (\bibinfo {year} {2019}),\ \bibinfo {note} {{DUNE}
  collaboration meeting}\BibitemShut {NoStop}%
\bibitem [{\citenamefont {Vilain}\ \emph {et~al.}(1994)\citenamefont {Vilain}
  \emph {et~al.}}]{Vilain:1994qy}%
  \BibitemOpen
  \bibfield  {author} {\bibinfo {author} {\bibfnamefont {P.}~\bibnamefont
  {Vilain}} \emph {et~al.} (\bibinfo {collaboration} {CHARM-II}),\ }\href
  {\doibase 10.1016/0370-2693(94)91421-4} {\bibfield  {journal} {\bibinfo
  {journal} {Phys. Lett.}\ }\textbf {\bibinfo {volume} {B335}},\ \bibinfo
  {pages} {246} (\bibinfo {year} {1994})}\BibitemShut {NoStop}%
\bibitem [{\citenamefont {Auerbach}\ \emph {et~al.}(2001)\citenamefont
  {Auerbach} \emph {et~al.}}]{Auerbach:2001wg}%
  \BibitemOpen
  \bibfield  {author} {\bibinfo {author} {\bibfnamefont {L.~B.}\ \bibnamefont
  {Auerbach}} \emph {et~al.} (\bibinfo {collaboration} {LSND}),\ }\href
  {\doibase 10.1103/PhysRevD.63.112001} {\bibfield  {journal} {\bibinfo
  {journal} {Phys. Rev.}\ }\textbf {\bibinfo {volume} {D63}},\ \bibinfo {pages}
  {112001} (\bibinfo {year} {2001})},\ \Eprint
  {http://arxiv.org/abs/hep-ex/0101039} {arXiv:hep-ex/0101039 [hep-ex]}
  \BibitemShut {NoStop}%
\bibitem [{\citenamefont {Deniz}\ \emph {et~al.}(2010)\citenamefont {Deniz}
  \emph {et~al.}}]{Deniz:2009mu}%
  \BibitemOpen
  \bibfield  {author} {\bibinfo {author} {\bibfnamefont {M.}~\bibnamefont
  {Deniz}} \emph {et~al.} (\bibinfo {collaboration} {TEXONO}),\ }\href
  {\doibase 10.1103/PhysRevD.81.072001} {\bibfield  {journal} {\bibinfo
  {journal} {Phys. Rev.}\ }\textbf {\bibinfo {volume} {D81}},\ \bibinfo {pages}
  {072001} (\bibinfo {year} {2010})},\ \Eprint {http://arxiv.org/abs/0911.1597}
  {arXiv:0911.1597 [hep-ex]} \BibitemShut {NoStop}%
\bibitem [{\citenamefont {Carena}\ \emph {et~al.}(2003)\citenamefont {Carena},
  \citenamefont {de~Gouv\^ea}, \citenamefont {Freitas},\ and\ \citenamefont
  {Schmitt}}]{Carena:2003aj}%
  \BibitemOpen
  \bibfield  {author} {\bibinfo {author} {\bibfnamefont {M.}~\bibnamefont
  {Carena}}, \bibinfo {author} {\bibfnamefont {A.}~\bibnamefont {de~Gouv\^ea}},
  \bibinfo {author} {\bibfnamefont {A.}~\bibnamefont {Freitas}}, \ and\
  \bibinfo {author} {\bibfnamefont {M.}~\bibnamefont {Schmitt}},\ }\href
  {\doibase 10.1103/PhysRevD.68.113007} {\bibfield  {journal} {\bibinfo
  {journal} {Phys. Rev.}\ }\textbf {\bibinfo {volume} {D68}},\ \bibinfo {pages}
  {113007} (\bibinfo {year} {2003})},\ \Eprint
  {http://arxiv.org/abs/hep-ph/0308053} {arXiv:hep-ph/0308053 [hep-ph]}
  \BibitemShut {NoStop}%
\bibitem [{\citenamefont {Tomalak}\ and\ \citenamefont
  {Hill}(2019)}]{Tomalak:2019ibg}%
  \BibitemOpen
  \bibfield  {author} {\bibinfo {author} {\bibfnamefont {O.}~\bibnamefont
  {Tomalak}}\ and\ \bibinfo {author} {\bibfnamefont {R.~J.}\ \bibnamefont
  {Hill}},\ }\href@noop {} {\  (\bibinfo {year} {2019})},\ \Eprint
  {http://arxiv.org/abs/1907.03379} {arXiv:1907.03379 [hep-ph]} \BibitemShut
  {NoStop}%
\bibitem [{\citenamefont {Hill}\ and\ \citenamefont
  {Tomalak}(2019)}]{Hill:2019xqk}%
  \BibitemOpen
  \bibfield  {author} {\bibinfo {author} {\bibfnamefont {R.~J.}\ \bibnamefont
  {Hill}}\ and\ \bibinfo {author} {\bibfnamefont {O.}~\bibnamefont {Tomalak}},\
  }\href@noop {} {\  (\bibinfo {year} {2019})},\ \Eprint
  {http://arxiv.org/abs/1911.01493} {arXiv:1911.01493 [hep-ph]} \BibitemShut
  {NoStop}%
\bibitem [{\citenamefont {Pickering}()}]{privatecomm}%
  \BibitemOpen
  \bibfield  {author} {\bibinfo {author} {\bibfnamefont {L.}~\bibnamefont
  {Pickering}},\ }\href@noop {} {}\bibinfo {note} {Private
  communication.}\BibitemShut {Stop}%
\bibitem [{\citenamefont {Golan}\ \emph {et~al.}(2012)\citenamefont {Golan},
  \citenamefont {Juszczak},\ and\ \citenamefont {Sobczyk}}]{Golan:2012wx}%
  \BibitemOpen
  \bibfield  {author} {\bibinfo {author} {\bibfnamefont {T.}~\bibnamefont
  {Golan}}, \bibinfo {author} {\bibfnamefont {C.}~\bibnamefont {Juszczak}}, \
  and\ \bibinfo {author} {\bibfnamefont {J.~T.}\ \bibnamefont {Sobczyk}},\
  }\href {\doibase 10.1103/PhysRevC.86.015505} {\bibfield  {journal} {\bibinfo
  {journal} {Phys. Rev. C}\ }\textbf {\bibinfo {volume} {86}},\ \bibinfo
  {pages} {015505} (\bibinfo {year} {2012})},\ \Eprint
  {http://arxiv.org/abs/1202.4197} {arXiv:1202.4197 [nucl-th]} \BibitemShut
  {NoStop}%
\bibitem [{\citenamefont {Magill}\ and\ \citenamefont
  {Plestid}(2017)}]{Magill:2016hgc}%
  \BibitemOpen
  \bibfield  {author} {\bibinfo {author} {\bibfnamefont {G.}~\bibnamefont
  {Magill}}\ and\ \bibinfo {author} {\bibfnamefont {R.}~\bibnamefont
  {Plestid}},\ }\href {\doibase 10.1103/PhysRevD.95.073004} {\bibfield
  {journal} {\bibinfo  {journal} {Phys. Rev.}\ }\textbf {\bibinfo {volume}
  {D95}},\ \bibinfo {pages} {073004} (\bibinfo {year} {2017})},\ \Eprint
  {http://arxiv.org/abs/1612.05642} {arXiv:1612.05642 [hep-ph]} \BibitemShut
  {NoStop}%
\bibitem [{\citenamefont {Ballett}\ \emph {et~al.}(2019)\citenamefont
  {Ballett}, \citenamefont {Hostert}, \citenamefont {Pascoli}, \citenamefont
  {Perez-Gonzalez}, \citenamefont {Tabrizi},\ and\ \citenamefont
  {Zukanovich~Funchal}}]{Ballett:2018uuc}%
  \BibitemOpen
  \bibfield  {author} {\bibinfo {author} {\bibfnamefont {P.}~\bibnamefont
  {Ballett}}, \bibinfo {author} {\bibfnamefont {M.}~\bibnamefont {Hostert}},
  \bibinfo {author} {\bibfnamefont {S.}~\bibnamefont {Pascoli}}, \bibinfo
  {author} {\bibfnamefont {Y.~F.}\ \bibnamefont {Perez-Gonzalez}}, \bibinfo
  {author} {\bibfnamefont {Z.}~\bibnamefont {Tabrizi}}, \ and\ \bibinfo
  {author} {\bibfnamefont {R.}~\bibnamefont {Zukanovich~Funchal}},\ }\href
  {\doibase 10.1007/JHEP01(2019)119} {\bibfield  {journal} {\bibinfo  {journal}
  {JHEP}\ }\textbf {\bibinfo {volume} {01}},\ \bibinfo {pages} {119} (\bibinfo
  {year} {2019})},\ \Eprint {http://arxiv.org/abs/1807.10973} {arXiv:1807.10973
  [hep-ph]} \BibitemShut {NoStop}%
\bibitem [{\citenamefont {Altmannshofer}\ \emph {et~al.}(2019)\citenamefont
  {Altmannshofer}, \citenamefont {Gori}, \citenamefont {Mart\'{i}n-Albo},
  \citenamefont {Sousa},\ and\ \citenamefont
  {Wallbank}}]{Altmannshofer:2019zhy}%
  \BibitemOpen
  \bibfield  {author} {\bibinfo {author} {\bibfnamefont {W.}~\bibnamefont
  {Altmannshofer}}, \bibinfo {author} {\bibfnamefont {S.}~\bibnamefont {Gori}},
  \bibinfo {author} {\bibfnamefont {J.}~\bibnamefont {Mart\'{i}n-Albo}},
  \bibinfo {author} {\bibfnamefont {A.}~\bibnamefont {Sousa}}, \ and\ \bibinfo
  {author} {\bibfnamefont {M.}~\bibnamefont {Wallbank}},\ }\href@noop {} {\
  (\bibinfo {year} {2019})},\ \Eprint {http://arxiv.org/abs/1902.06765}
  {arXiv:1902.06765 [hep-ph]} \BibitemShut {NoStop}%
\bibitem [{\citenamefont {Abgrall}\ \emph {et~al.}(2011)\citenamefont {Abgrall}
  \emph {et~al.}}]{Abgrall:2011ae}%
  \BibitemOpen
  \bibfield  {author} {\bibinfo {author} {\bibfnamefont {N.}~\bibnamefont
  {Abgrall}} \emph {et~al.} (\bibinfo {collaboration} {NA61/SHINE}),\ }\href
  {\doibase 10.1103/PhysRevC.84.034604} {\bibfield  {journal} {\bibinfo
  {journal} {Phys. Rev. C}\ }\textbf {\bibinfo {volume} {84}},\ \bibinfo
  {pages} {034604} (\bibinfo {year} {2011})},\ \Eprint
  {http://arxiv.org/abs/1102.0983} {arXiv:1102.0983 [hep-ex]} \BibitemShut
  {NoStop}%
\bibitem [{\citenamefont {Akaishi}\ \emph {et~al.}(2019)\citenamefont {Akaishi}
  \emph {et~al.}}]{Akaishi:2019dej}%
  \BibitemOpen
  \bibfield  {author} {\bibinfo {author} {\bibfnamefont {T.}~\bibnamefont
  {Akaishi}} \emph {et~al.} (\bibinfo {collaboration} {EMPHATIC}),\ }\href@noop
  {} {\  (\bibinfo {year} {2019})},\ \Eprint {http://arxiv.org/abs/1912.08841}
  {arXiv:1912.08841 [hep-ex]} \BibitemShut {NoStop}%
\end{thebibliography}%
